\RequirePackage{lineno}

\documentclass[aps, pra, 10pt, superscriptaddress, twocolumn, nofootinbib]{revtex4-1}
\usepackage[UKenglish]{babel}
\usepackage{latexsym}
\usepackage[pdftex, colorlinks=true, linkcolor=red, citecolor=blue, urlcolor=magenta, pdfstartview=FitH]{hyperref}
\usepackage{graphicx}	
\usepackage{amssymb}
\usepackage{amsmath}
\usepackage{xcolor}

\usepackage[normalem]{ulem}

\begin{document}


\title{Dimensional crossover of acoustic phonon lifetime in $2H$-MoSe$_2$}

\author{P. Soubelet}
\affiliation{Centro At\'omico Bariloche \& Instituto Balseiro (C.N.E.A.) and CONICET, 8400 S.C. de Bariloche, R.N., Argentina.}

\author{A.~A. Reynoso}
\affiliation{Centro At\'omico Bariloche \& Instituto Balseiro (C.N.E.A.) and CONICET, 8400 S.C. de Bariloche, R.N., Argentina.}

\author{A. Fainstein}
\affiliation{Centro At\'omico Bariloche \& Instituto Balseiro (C.N.E.A.) and CONICET, 8400 S.C. de Bariloche, R.N., Argentina.}

\author{K. Nogajewski}
\affiliation{Institute of Experimental Physics, Faculty of Physics, University of Warsaw, Pasteura 5, 02-093 Warsaw, Poland.}
\affiliation{Laboratoire National des Champs Magn\'etiques Intenses (CNRS, UJF, UPS, INSA), BP 166, 38042 Grenoble Cedex 9, France.}

\author{M. Potemski}
\affiliation{Institute of Experimental Physics, Faculty of Physics, University of Warsaw, Pasteura 5, 02-093 Warsaw, Poland.}
\affiliation{Laboratoire National des Champs Magn\'etiques Intenses (CNRS, UJF, UPS, INSA), BP 166, 38042 Grenoble Cedex 9, France.}

\author{C. Faugeras}
\email[E-mail: ]{clement.faugeras@lncmi.cnrs.fr}
\affiliation{Laboratoire National des Champs Magn\'etiques Intenses (CNRS, UJF, UPS, INSA), BP 166, 38042 Grenoble Cedex 9, France.}

\author{A.~E. Bruchhausen}
\email[E-mail: ]{Axel.Bruchhausen@cab.cnea.gov.ar}
\affiliation{Centro At\'omico Bariloche \& Instituto Balseiro (C.N.E.A.) and CONICET, 8400 S.C. de Bariloche, R.N., Argentina.}

\date{\small \today}

\begin{abstract}
A time-resolved observation of coherent interlayer longitudinal acoustic phonons in 2$H$-MoSe$_2$ is reported. A femtosecond pump-probe technique is used to investigate the evolution of the energy loss of these vibrational modes in a wide selection of MoSe$_2$ flakes with different thicknesses ranging from bilayer up to the bulk limit. By directly analysing the temporal decay of the modes, we can clearly distinguish an abrupt crossover related to the acoustic mean free path of the phonons in a layered system, and the constraints imposed to the acoustic decay channels when reducing the dimensionality. Loses intrinsic to the low dimensionality of single or few layer materials impose critical limitations for their use in optomechanical and optoelectronic devices.
\end{abstract}

\maketitle

\section{Introduction \& motivation}

Two dimensional (2D) transition metal dichalcogenides (TMDCs) \cite{Wang-NatNanotechnol7-699(12)} combine many different unique properties such as high in-plane mobility \cite{Ayari-JournalOfAppliedPhysics-101(07), Nam-ScientificReports5-10546(15)}, relatively high heat conduction \cite{PengRSC6-5767(16), Faugeras-ACSNano8-1889(10)}, large Seebeck coefficients \cite{Buscema-NanoLetter13-2(13)}, significant spin-orbit coupling \cite{Wang-NatNanotechnol7-699(12)}, together with remarkable mechanical properties \cite{CastellanosGomez-AdvancedMaterials24-772(12), Cooper-PRB87-035423(13), CastellanosGomez-AdvancedMaterials25-6719(13)}. 
Some members of this $MX_2$ family, with $2H$ phase, and where $M$ is W or Mo and $X$ is S, Se or Te, are semiconductors with relatively large indirect band gaps. In the form of monolayers, they become direct band gap semiconductors with strong excitonic effects and strong light-matter coupling at room temperature due to the reduced dimensionality \cite{Wang-RevModPhys90-021001(18), Splendiani-NanoLetters10-4(10), Mak-PRL105-136805(10), Mak-NatureMaterials12-207(12)}. They provide a strong luminescence, and they represent an interesting complement to gapless graphene mainly as photoactive materials in the NIR-Vis range.
Applications seeking the conception of optoelectronic devices based on these TMDCs nanomaterials \cite{Back-PRL120-037401(18), Scuri-PRL120-037402(18)}, ultrafast photodetection and light emission \cite{Wang-NatNanotechnol7-699(12), Huang-Nanotechnology27-445201(16), Bie-NatureNanotechnology12-1124(17)}, valleytronics and spintonics \cite{Schaibley-NatRevMater1-16055(16), Mak-NatureNanotechnology7-494(12), Xiao-PRL108-196802(12), Xu-NaturePhysics10-343(14)}, field effect transistors (FETs) \cite{Wang-NatNanotechnol7-699(12), Nam-ScientificReports5-10546(15)} based on few layer van der Waals heterojunctions \cite{Huang-NatureNanotechnology12-1148(17), Larentis-APL101-223104(12)}, represent just a glimpse of the recent fruitful activity in the field.

Similar to graphene, TMDC monolayers constitute an uttermost 2D crystalline system composed of atoms linked by strong covalent bonds. In analogy to graphite, in a bulk TMDC crystal these 2D-TMDC layers are stacked one on top of another and held together by ``weak'' van der Waals type interactions \cite{Geim-Nature499-419(13), Ribero-PRB90-115438(14)}, and exfoliation out from ultra-pure bulk crystals is one technique that allows to isolate mono- and also few layers of these materials. MoSe$_2$ is the prototype of 2D semiconductor with bright exciton ground state \cite{Molas-2DMaterials4-021003(17)} and a well defined emission spectrum \cite{Arora-Nanoscale7-20769(15)}.

As in other materials, lattice vibrations (phonons) of these TMDCs play an essential role in determining their physical properties \cite{Ghosh-NatureMaterials9-555(10), PengRSC6-5767(16), Wang-NatNanotechnol7-699(12)}, and mono- and few-layers of different TMDCs, especially semiconducting ones, have recently been subject of intense investigations \cite{Lu-NanoResearch9-3559(16), MolinaSanchez-SurfSciRep70-554(15), Soubelet-PRB93-155407(16), Molina-SanchezPhysRevB.84.155413(11), Horzum-PRB87-125415(13), Sekine-JPSJournal49-3(80), Tongay-NanoLetters12-11(12), Tonndorf-OptExpress21-4(13), Kumar-Nanoscale6-9(14), Lin-NatureCommunications8-1745(17)}. 
It is worth noting that layered crystals constitute a natural van der Waals structure \cite{Geim-Nature499-419(13)}, and are hence an ideal system for probing interlayer vibrational modes and the cross-plane forces stand behind of them \cite{Zhao-NanoLetters13-1007(13), Zhang-PRB87-115413(13)}. 
One of the most elusive quantities in the studies of lattice vibrations has so far been the lifetime of phonons, in general one of the least known properties of solid state systems. The reason for this should probably be ascribed to experimental challenges associated with direct quantitative characterization of processes leading to the phonon decay and coherence loss, as well as the complexity of modelling them \cite{Daly-PRB70-21(04), Daly-PRB80-17(08), Bruchhausen-PRL106-077401(11), Cuffe-PRL110-9(13), Maris-PhysicalAcoustics-Book(71)}.\\
\begin{figure*}[ttt!!]
\includegraphics*[keepaspectratio=true, clip=true, angle=0, width=1.3\columnwidth, trim={19mm, 14mm, 25mm, 18mm}]{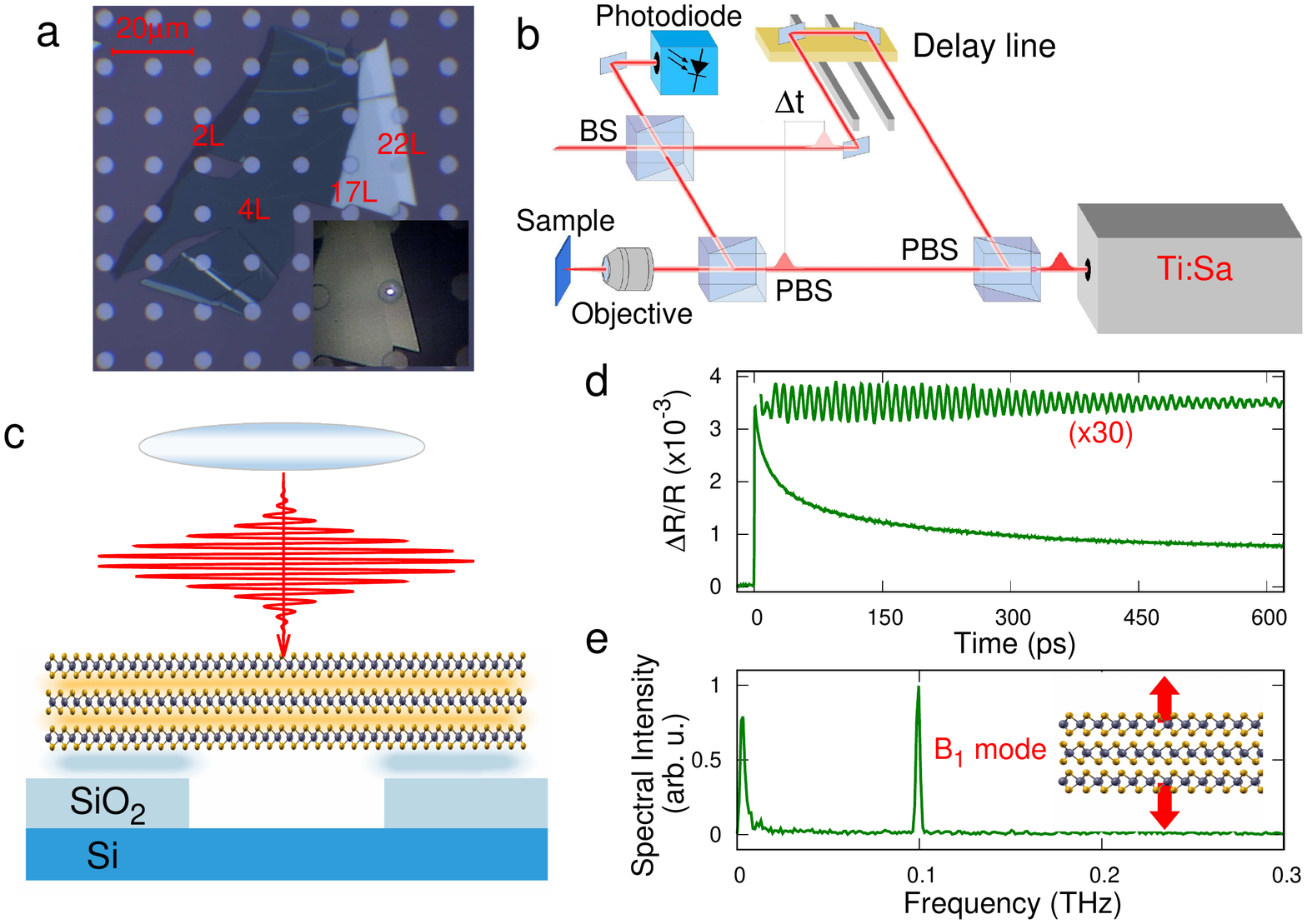}
\caption{\textbf{Samples and experimental set up. a)} Optical image taken with a 50$\times$ objective of an exfoliated flake. The color contrast outside the patterned holes enables the identification of sample thicknesses. The inset shows part of this image acquired with the 100$\times$ objective, where the laser spot is observed centred on top of the patterned hole of a samples with 22L. \textbf{b)}  Schematic diagram of the pump-probe setup. \textbf{c)} Sketch of the incident laser pulse on a free-standing 3L MoSe$_2$ flake on top of a patterned hole. The shaded gaps inbetween MoSe$_2$ layers (yellow) and between MoSe$_2$ and SiO$_2$ (light-blue) indicate different van der Waals interactions. \textbf{d)} Typically obtained transient reflectivity $\Delta R/R$ for 22L. The filtered signal displaying the coherent acoustic $B_1$ oscillations is shown on top ($\times 30$), and the corresponding numerical Fourier transform is depicted in \textbf{e)}. The inset in panel e) exemplifies the dynamics of the $B_1$ mode for a 3L system.}
\label{figura:1}
\end{figure*}
Since TMDCs exhibit strong correlations between electronic states and lattice vibrations \cite{Lin-NatureCommunications8-1745(17), Mannebach-NanoLetters17-7761(17)}, which naturally affect a whole range of fundamental properties of these materials, e.g. thermal transport, carrier mobility, light emission, among others \cite{Wang-NatNanotechnol7-699(12), Ayari-JournalOfAppliedPhysics-101(07), Nam-ScientificReports5-10546(15), PengRSC6-5767(16), Faugeras-ACSNano8-1889(10), Buscema-NanoLetter13-2(13), CastellanosGomez-AdvancedMaterials24-772(12), Cooper-PRB87-035423(13), CastellanosGomez-AdvancedMaterials25-6719(13), Wang-RevModPhys90-021001(18), Splendiani-NanoLetters10-4(10), Mak-PRL105-136805(10)}, having a good characterization of the phonon modes together with their damping (decay) rates is thus crucial to understand the possible decoherence channels that exist in these 2D-TMDCs nanostructures, and are hence essential for the conception of electronic and optoelectronic devices \cite{Morell-NanoLetters16-5102(16), Guettinger-NatureNanotechnology12-631(17), Scuri-PRL120-037402(18), Back-PRL120-037401(18)}.

Only a very few recent investigations have dealt with direct time-domain analysis of the actual dynamics of the electronic \cite{Shi-ACSNano7-1072(13), Wang-ACSNano7-11087(13), Mannebach-ACSNano8-107324(14), He-Nanoscale9-9526(15), Czech-ACSNano9-12146(15), Ceballos-AdvancedFunctionalMaterials27-1604509(17)} and phononic \cite{Boschetto-NanoLetters13-4620(13), Ge-ScientificReports4-5722(14), Jeong-ACSNano10-5560(16), Beardsely-ScientificReports6-26970(16), He-ScientificReports6-30487(16), Matis-ScientificReposts7-5656(17), Kim-APLMaterials5-086105(17), Lin-NatureCommunications8-1745(17)} modes in TMDCs. In this paper, we focus on the temporal dynamics of a very particular acoustic phonon mode in $2H$-MoSe$_2$, which tests the very nature of the interlayer forces: The acoustic interlayer breathing mode ($B_1$) \cite{Froehlicher-NanoLetters15-6481(2015), Ji-PhysicaE80-130(16), Liang-ACSNano11-11777(17)}.  
By tracing the phonon dynamics directly in the time-domain as function of the number of MoSe$_2$ layers, covering a wide range of number of layers, varying from $2,3,\dots$ individual layers up to a ``bulk'' situation, we are able to clearly evidence the \emph{dimensional crossover} of the lifetime of the breathing mode ($\tau_{B_1}$). Indirect effects on the material properties due to a dimensional crossover on these kind of atomic layered materials, where the nature of the phonon scattering plays a decisive role, have been observed mainly in thermal transport and heat conduction \cite{Ghosh-NatureMaterials9-555(10), Yang-PRE74-062101(06), Gu-ArXiv-1705.06156v1(17)}, but the direct measurement of the phonon lifetime for systems with different number of layers, presented in this study, was still lacking.

\section{Results}
\paragraph*{\bf Samples and characterization:}
In order to isolate the crystal from possible interfering contact effects with the substrate \cite{Ji-PhysicaE80-130(16), Buscema-NanoResearch7-561(14), Lu-NanoResearch9-3559(16)}, the exfoliated 2$H$-MoSe$_2$ flakes were deposited on a specially designed substrate, which consisted on a (001) Si wafer with a 90\,nm SiO$_2$ layer formed on-top of it, and where regular 6\,$\mu$m diameter circular holes were patterned in the SiO$_2$. The flakes were randomly scattered on the substrate, and a careful selection enabled the identification of different flakes that were found free-standing on the holes, i.e. with no contact of the flakes' surfaces with the substrate. Figure\,\ref{figura:1}a displays an optical image with an example of several flakes scattered on the patterned substrate. We have identified the number of layers of the different flakes by their distinct optical contrast and with Raman scattering spectroscopy \cite{Soubelet-PRB93-155407(16), SOM}. In Fig.\ref{figura:1}a, the regions are labeled with the identified number of layers. The limiting border between different layer number, and the superposition with the underlying holes can be clearly observed.

It is important to mention that the low frequency geometric ``drum-like'' oscillations formed by the suspended region cannot be accessed with our experiment \cite{Morell-NanoLetters16-5102(16), Will-NanoLetters17-5950(17), Guettinger-NatureNanotechnology12-631(17)}. Our aim here is to analyse the \textit{isolated} flakes, to be able to have a precise determination of the intrinsic temporal development of the  internal ``breathing'' $B_1$ modes of bi- and few-layer membranes. When the samples are supported, the boundary conditions are changed considerably, modifying the acoustic dynamics and the time-domain signal.\\  

\paragraph*{\bf Pump-probe spectroscopy:}
Figure\,\ref{figura:1}b shows a typical pump-probe set-up, such as the one used in this work. The 100\,fs laser pulses are provided by a Ti:Sapphire oscillator with an 80\,MHz repetition rate and a central wavelength of about 805\,nm. The laser was split in two, a first more intense part (pump-beam) was focused directly onto the sample. The weaker part (probe-beam) was time-delayed with respect to the pump-beam using a mechanical delay-line, and focused spatially superimposed to the pump-beam onto the sample. All experiments were carried out at room temperature. Both beams, pump and probe, were focused co-linearly through the same 100$\times$ microscope objective (NA=0.95), that could be simultaneously used to acquire a white light image (see inset of Fig.\ref{figura:1}a). The spot size was of $\varnothing\sim 1\,\mu$m and enabled the precise addressing of each individual free-standing MoSe$_2$ flake within one of the patterned holes, as is shown in the inset of Fig.\ref{figura:1}a. Figure\,\ref{figura:1}c sketches this situation.

The temporal modulation of the reflection of the probe-beam $\Delta R(t)$, due to the changes in the optical constants by the impulsive excitation of vibrations induced by the pump-beam, was measured synchronously using a lock-in amplifier and a photo-diode. Figure\,\ref{figura:1}d shows a typical ``as measured'' transient obtained for a 22-layered (22L) flake, displaying a strong onset when both, pump and probe beams, coincide temporally ($t=0$: zero delay-time), and relaxing multi-exponentially to its equilibrium. This behaviour mainly reflects the contribution of the electronic dynamics after the pump excitation to the temporal modulation of the optical constants. The signal on top ($\times$30) shows the extracted high-frequency oscillation modes corresponding to the interlayer vibrational breathing $B_1$ mode, which rings down with a characteristic damping time. It's numerical Fourier transform (nFT) is shown in Fig.\ref{figura:1}e displaying the clear single peak with a frequency of about 0.1\,THz. The inset sketches the $B_1$ optically active mode for a MoSe$_2$ system consisting of three layers (3L). Here the two outer layers move in the opposite direction, indicated by the arrows \cite{Zhao-NanoLetters13-1007(13), Lu-NanoResearch9-3559(16), Zhao-PRB90-245428(14)}. 
\begin{figure}[ttt!!]
\includegraphics*[keepaspectratio=true, clip=true, angle=0, width=\columnwidth, trim={17mm, 14mm, 26mm, 17mm}]{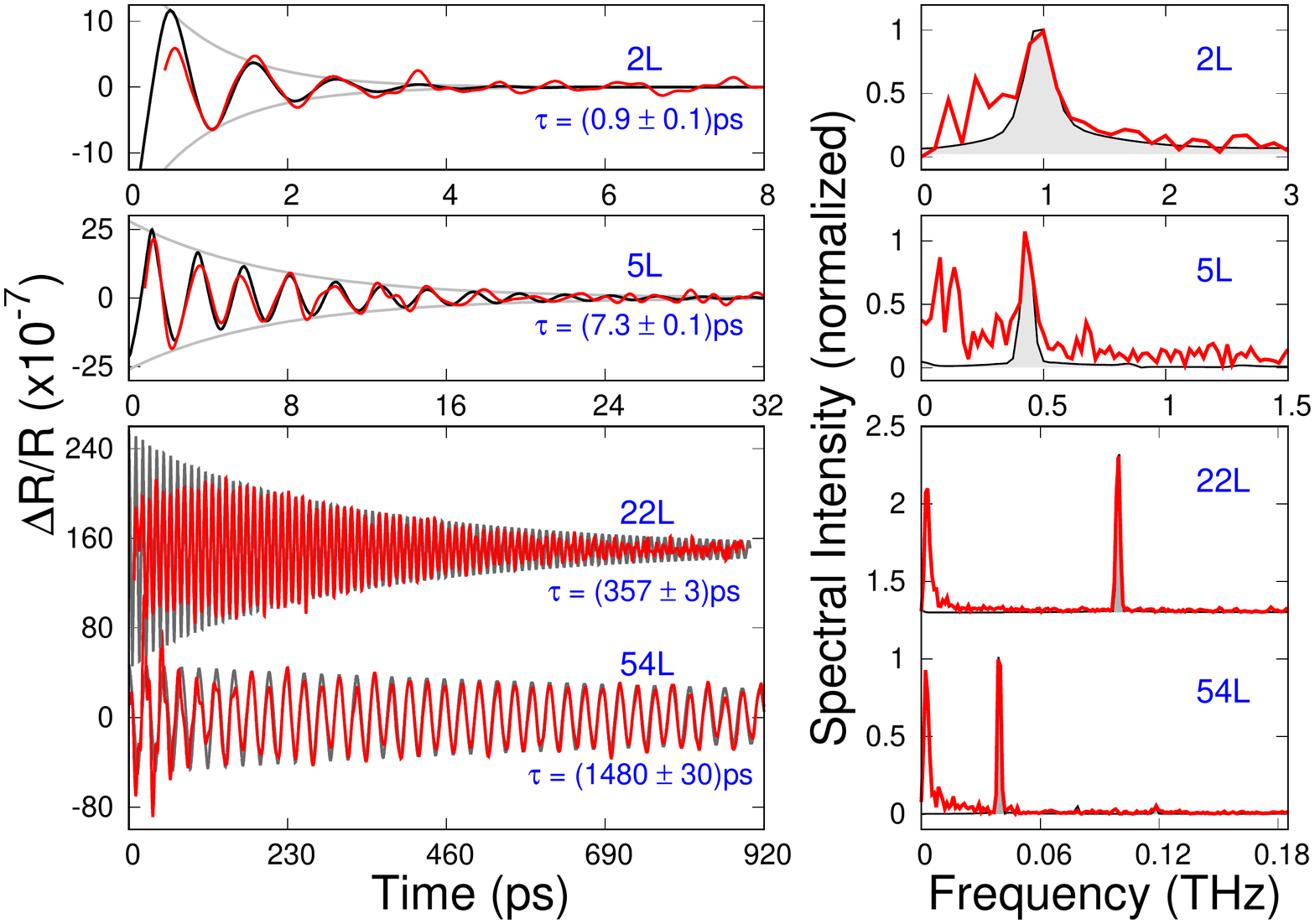}
\caption{\textbf{Temporal and Fourier domain of the measured transients}. Left: Experimental signals (red curves) in the time-domain showing the extracted oscillations corresponding to the $B_1$ mode for samples with different layer number (as indicated). Right: Numerical Fourier Transform of the corresponding transients. The black curves are the simulations that best fit the experiments simultaneously in time and frequency domains. The obtained damping times $\tau$ are shown below each transient, and the corresponding decaying exponential envelopes are indicated in grey.}
\label{figura:2}
\end{figure}
The transient reflectivity pump-probe measurements have been performed on more than 20 free-standing MoSe$_2$ flakes, using typical mean powers of 400\,$\mu$W and 100\,$\mu$W for the pump and for probe, respectively. In Fig.\ref{figura:2}, examples of the extracted breathing mode oscillations, for samples with different number of layers, are displayed with red lines. The panels on the left side correspond to the coherently excited longitudinal acoustic oscillations in the time-domain, whereas the panels on the right side show the corresponding nFT of the oscillations. Note the strong change of the $B_1$ modes frequency and linewidth with the number of MoSe$_2$ layers. The fact that the oscillations damping time is significantly shorter for samples with less number of layers, is systematic and central for this work. This is also noticeable for the nFTs, where for the thicker samples (22L and 54L) the spectral width of the peaks is Fourier limited by the temporal window of observation. The frequency of the B$_1$-modes can be very well established (Fig.\ref{figura:3}), and we can see that it can be as high as $\sim$1\,THz for the thinnest possible sample with two layers (2L), and shifting down when the stacking number increases.\\
\begin{figure}[!t]
\includegraphics*[keepaspectratio=true, clip=true, angle=0, width=\columnwidth, trim={18mm, 14mm, 29mm, 20mm}]{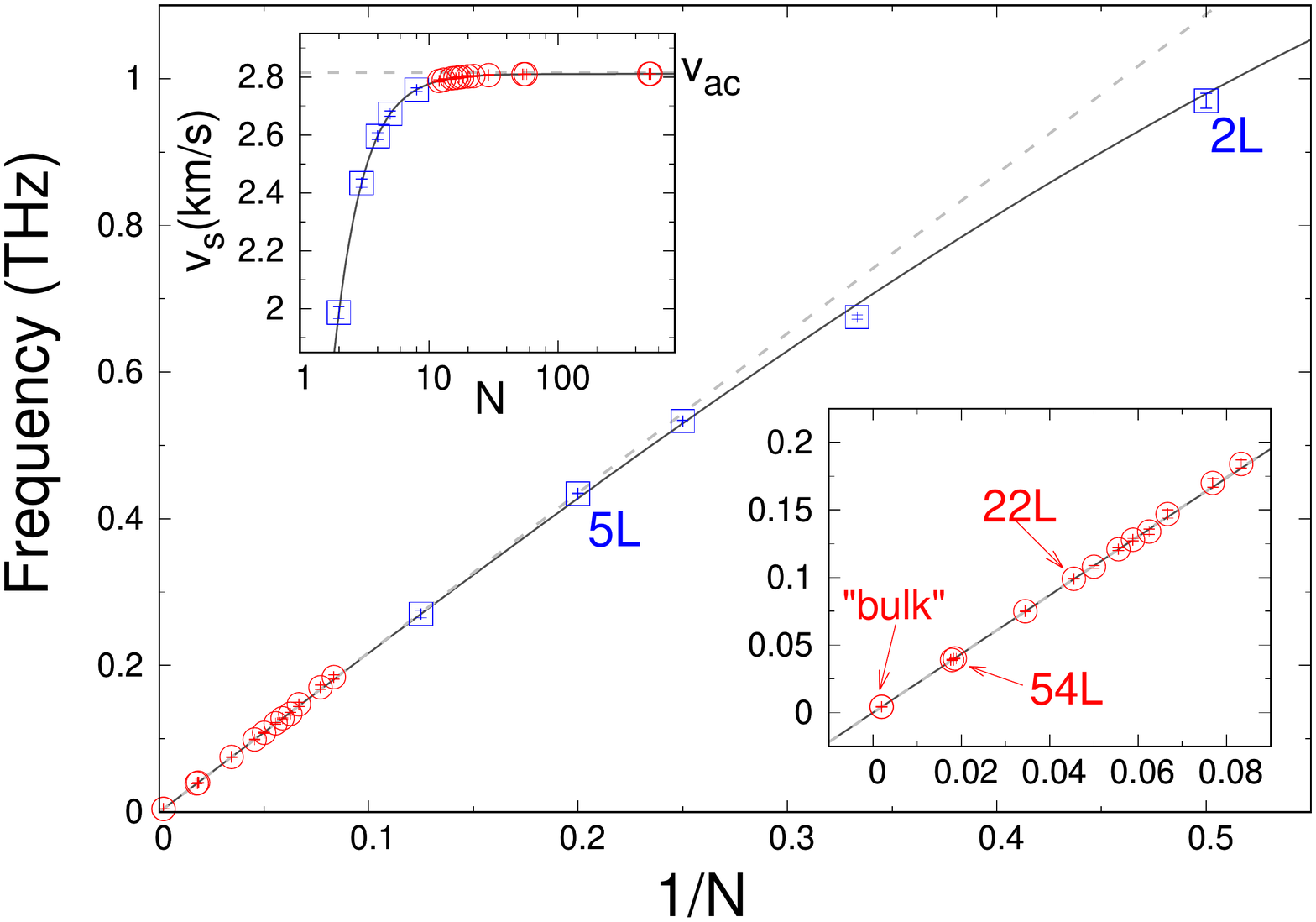}
\caption{\textbf{Modes frequency as a function of the inverse number of layers $1/N$}. The blue squares correspond to the thinner samples in which the number of layers was experimentally identified. The full curve is the fitting with the linear chain model, and the red circles corresponds to the thicker samples, where the values of $N$ were determined by using the obtained value for $f_o$. The inset on the bottom right shows a detail of these latter cases. The top-left inset corresponds to the calculated group velocity for the measured points, and the full curve to the calculated curve as a function of $N$ (see the text for details regarding the model). The dashed lines indicate the result obtained for a continuum elastic model with speed of sound $v_{ac}$.}
\label{figura:3}
\end{figure}

\paragraph*{\bf Linear chain model:}
The $B_1$ longitudinal acoustic breathing phonon modes have been observed by Raman spectroscopy \cite{Zhao-NanoLetters13-1007(13), Lu-NanoResearch9-3559(16), Zhao-PRB90-245428(14), Froehlicher-NanoLetters15-6481(2015), Soubelet-PRB93-155407(16)}, and are characterized by compressing-expanding the different layers against each other, but leaving the internal structure within each individual layer intact (see the sketch in inset of Fig.\ref{figura:1}d), i.e. affecting only the weak interlayer forces. It has been proven for similar systems that a linear chain model with effective masses for each layer (per unit area, $\mu$), and an effective interlayer elastic force (with elastic constant per unit area, $K$) is well suited to describe these modes \cite{Froehlicher-NanoLetters15-6481(2015), Ji-PhysicaE80-130(16), Liang-ACSNano11-11777(17)}. To solve the elastic equation of motion for the unsupported flakes, free-surface (stress-free) boundary conditions are proposed, resulting in the following well-known dispersion relation for the frequency of the modes as a function of the number of layers ($N$) \cite{Ji-PhysicaE80-130(16), Liang-Nanoscale9-15340(17), Liang-ACSNano11-11777(17)} 
\begin{eqnarray}\label{eqn-frequency linear chain model}
f_{N,n}=f_o\sin\big(\mbox{$\frac{k_{N,n}\,d_o}{2}$}\big)\ ,
\end{eqnarray}
where $d_o=6.459$\,\AA\,\cite{Roy-ACS-AppMat8-11(16), Coehoorn-PRB35-12(87)} is the interlayer distance, $k_{N,n}=\frac{2\pi}{\lambda_{ac}}$, the associated acoustic wavelength is $\lambda_{ac}=\frac{2 N d_o}{n}$, $f_o$ is related to the interlayer elastic constants and the effective mass as
\begin{eqnarray} \label{eqn - f_o}
f_o=\sqrt{\mbox{$\frac{K}{\pi^2 \mu}$}}\ ,
\end{eqnarray}
and $n=1, 2, \dots (N-1)$ corresponds to the modes index. For the purpose of the present investigation we only consider the fundamental mode $n=1$. For the thinner samples ($N$=2, 3, 4, 5 and 8 layers), the number of layers can be well identified, and their obtained frequency is plotted in Fig.\ref{figura:3} with the blue squares. These results are fitted with the above expression \eqref{eqn-frequency linear chain model} obtaining a value for $f_o=(1.39\pm 0.03)$\,THz. The fitted curve is shown in Fig.\ref{figura:3} with the full curve. For thicker samples, the values of $N$ can be determined by using the obtained value for $f_o$, the modes frequency obtained from the Fourier analysis (see e.g. Fig.\ref{figura:2}, right panels), and deriving $N$ from eqn.\eqref{eqn-frequency linear chain model}. The results are plotted with red circles in Fig.\ref{figura:3}. The inset on the right shows a close up for these cases. For the limiting case of large $N$ the linear dependence is obtained. For thinner samples a slight bending of the full line can be noticed, indicative of a sound speed reduction (see top-left inset in Fig.\ref{figura:3}). The overall agreement is very good. The dashed grey line corresponds to the case of a linear dispersion obtained using the continuum elastic model.  The samples labelled as ``bulk'' actually corresponds to the largest $N$ found and was estimated to be of 519$\pm$5 layers. An estimation for the in-plain effective mass for each layer, considering the atomic masses and the MoSe$_2$ in-plane unit cell \cite{Roy-ACS-AppMat8-11(16), Coehoorn-PRB35-12(87)}, gives $\mu\simeq 4.41\times 10^{-6}$\,kg/m$^2$. From $f_o$, obtained from the above fit, we can derive using eqn.\eqref{eqn - f_o} an effective interlayer elastic force constant $K\simeq 8.42\times 10^{18}$\,N/m$^2$, consistent with other van der Waals materials \cite{Zhao-NanoLetters13-1007(13), Froehlicher-NanoLetters15-6481(2015)}.\\

Given the above dispersion relation \eqref{eqn-frequency linear chain model}, it is possible to derive the longitudinal acoustic propagation velocity $v_{s}$ in the stacking direction. This group velocity, defined as $\frac{d\omega}{dk}$ ($\omega=2\pi f_{N}$), gives:
\begin{eqnarray}\label{eqn-speed of sound linear chain model}
v_{s}(N)= \pi\,f_o\,d_o\,\cos\big(\mbox{$\frac{k_{N}\,d_o}{2}$}\big)\ .
\end{eqnarray}
The upper left inset in Fig.\ref{figura:3} shows the calculated velocity for the corresponding points displayed in the main figure. The dashed curve in this inset, corresponds to the interpolated calculated curve as a function of $N$. The asymptotic value for the ``bulk'' situation (large $N$), yields a ``bulk'' longitudinal acoustic sound velocity of $v_{ac}=\pi f_o d_o=2820$\,m/s. Notice that the low value of this velocity is compatible with the weak coupling between layers, and is similar to values obtained for similar 2D TMDCs \cite{Liang-ACSNano11-11777(17), Ge-ScientificReports4-5722(14)}.\\

\paragraph*{\bf Simulations:}
In order to gain a better understanding of the physical processes responsible for the generation and detection of these coherent longitudinal acoustic phonons and the resulting shape of the transient modulation of the probes reflectivity $\frac{\Delta R(t)}{R}$, we have modelled the complete acoustic impulsive generation and detection processes. The modelling considers the propagation and the modification of the electromagnetic fields within the MoSe$_2$ membrane for the pump and the probe pulses  \cite{Matsuda-JOptSocAmB19-3028(02), PascualWinter-PRB85-235443(12)} and adapting the theory to include the elastic acoustic part accounting for the modes resulting from the linear chain model. The impulsive absorption of the pump pulse and the consequent phonon generation is described considering the displacive electro-optic mechanism \cite{PascualWinter-PRB85-235443(12)}, whereas the coupling of the electromagnetic probe pulse and the phonons assumes a photo-elastic process \cite{PascualWinter-PRB85-235443(12)}, i.e. longitudinal acoustic phonons modulate the dielectric susceptibility through the generated acoustic interlayer strain within the free-standing flake. In order to describe the temporal decay of the observed oscillations, we have introduced a dissipative term to the linear chain that adds the additional damping constant ($\tau^{-1}$), where $\tau$ represents the acoustic damping time. 

\begin{figure}[!t]
\includegraphics*[keepaspectratio=true, clip=true, angle=0, width=\columnwidth, trim={29mm, 14mm, 42mm, 17mm}]{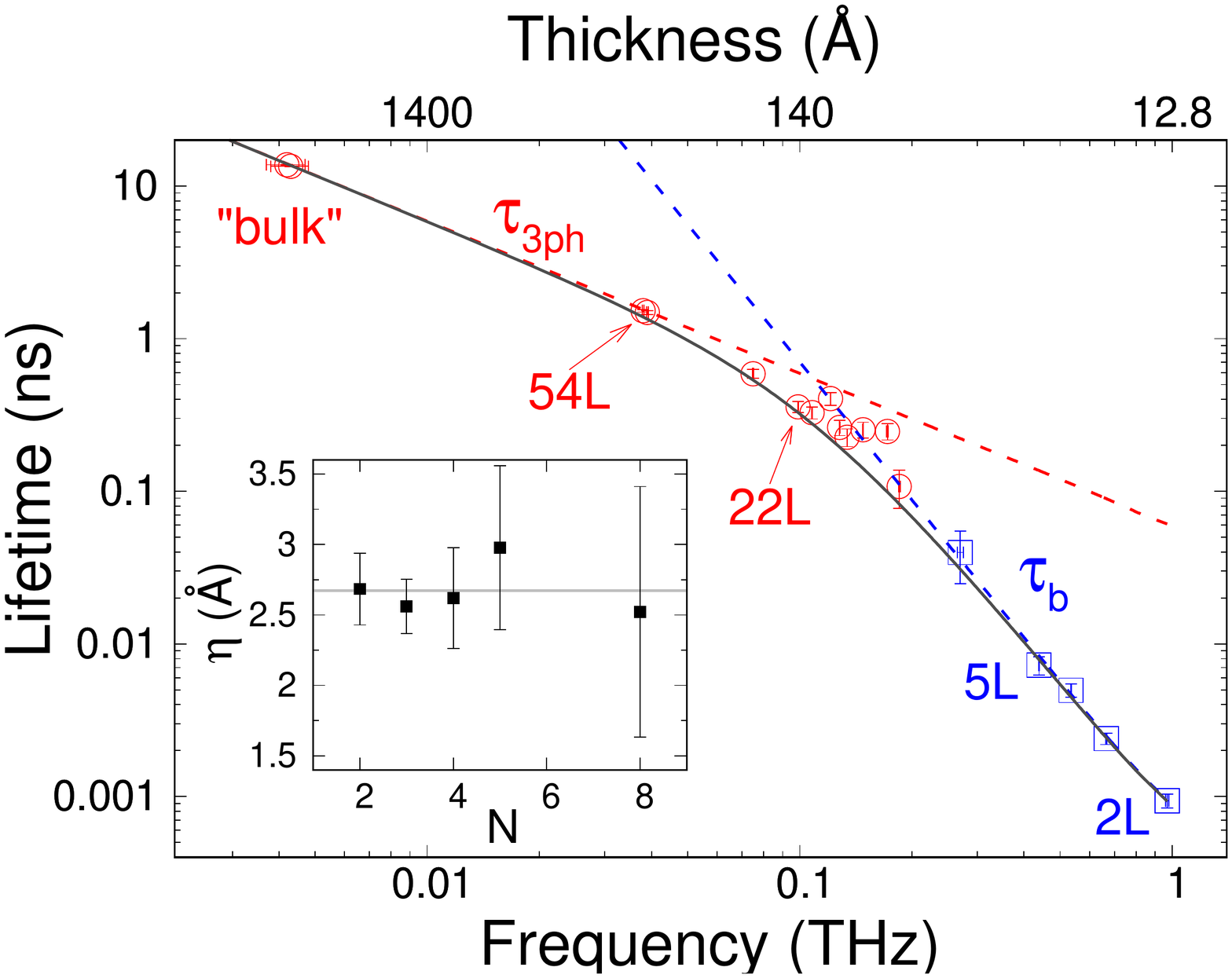}
\caption{\textbf{Phonon lifetime $\tau$ as function of the corresponding $B_1$-mode frequency}. Those samples in which $N$ was experimentally identified are plotted with blue squares and those in which it was determined by means of the fitted $f_o$ are plotted with red circles. The simulation obtained using the three-phonon scattering model is plotted with the red dashed line, and in blue the simulation accounting the boundary scattering mechanism. The total contribution to the modes lifetime can be estimated using the Matthiesen's rule and it is plotted with the grey line. The inset displays the asperity values obtained for each of the blue squares and the average is indicated by the horizontal line.}
\label{figura:4}
\end{figure}

The results of the simulations, given the obtained values for $N$, the modes frequency $f_N$, and the bulk index of refraction for MoSe$_2$ \cite{Soubelet-PRB93-155407(16)}, basically leaves two \textit{a priori} uncorrelated parameters to adjust: First, the photo-elastic constant, which accounts only for a multiplicative constant \cite{PascualWinter-PRB85-235443(12)}; and second, the damping time ($\tau$). In Fig.\ref{figura:2}, for the four samples, we exemplify how the simulations (black curves) fit the measured data. The agreement is quite remarkable for all cases in both domains, the temporal (left panels) as well as the spectral (right panels). The corresponding $\tau$ is indicated together with the used $N$. It is worth mentioning that both values are extremely critical for determining the central frequency and the correct \emph{simultaneous} adjustment of the temporal traces and the spectral domain. Changing $N$ in $\pm$ 1 layer, or modifying $\tau$ slightly, worsens the adjustment rapidly. The acoustic lifetimes are central to this work and of major interest for applications. In Fig.\ref{figura:4} we plot the obtained lifetimes $\tau$ for each of the measured samples, as a function of the corresponding frequency of the $B_1$-mode.

designed the samples; D.G.S. and A.S. grew and char-
acterized the samples
The general behaviour is that $\tau$ decreases for increasing phonon frequency, i.e. when the number of layers decreases. The evolution is rather linear  for low and high frequencies (within the log-log scale), but a major and significant change in the slope above $\sim$0.1\,THz (i.e. below $\sim$20 layers) can be clearly observed. Lower frequencies have a dependence that is proportional to $f^{-1}$, while the higher frequencies are better described by a curve $\propto f^{-3}$. This behaviour strongly suggests a fundamental change in the regime responsible for the energy loss of the observed longitudinal acoustic breathing $B_1$ modes, when lowering the number of layers.

\section{Discussion}\label{sec: Discussion}

The intrinsic lifetime of propagating acoustic phonons in ultra-pure  bulk matter is mainly determined by anharmonicity, i.e. the interaction of the coherently generated acoustic modes with the existing thermal phonon bath, through three-phonon-scattering processes \cite{Maris-PhysicalAcoustics-Book(71)}.
Several methods have been proposed to calculate the phonon lifetime, but due to the complex nature of the different phonon-phonon interactions contributing to the decay channels, and despite the importance and technological interest, a complete general and rigorous modeling has been rather elusive. Several limiting cases have been treated, depending on the different regimes of relative temperature and acoustic frequencies \cite{Maris-PhysicalAcoustics-Book(71), Srivastava-Book(90)}. At a given temperature, $\tau^{-1}$ generally shows a polynomial-like dependence with the frequency ($\tau^{-1}\propto f^m$) \cite{Srivastava-Book(90), AlOtaibi-JOfPhys-CondMatt27-335801(15)}. 

Given the fact that we observe a $f^{-1}$ dependence for $f\lesssim 0.1$\,THz, we choose a model based on the linearised Boltzmann equation in combination with a first order time-dependent perturbation theory to the anharmonic potential to account for the phonon-phonon scattering rates \cite{Srivastava-Book(90)}. In particular, the three-phonon scattering rate is approximated using the single-mode relaxation time (SMRT) approximation, assuming a simple Debye model adapted for anisotropic materials \citep{Chen-PRB.87.125426(13)}, and only including interactions of the observed $B_1$ modes with acoustic phonon branches of the membranes. The polarization of the observed modes is longitudinal ($L$). Consequently we will \textit{a priori} be considering three-phonon scattering (normal and umklapp) processes of the type 
\begin{eqnarray}
\omega_{L}+\omega_{s'}\rightarrow\omega_{s''}\text{~ ~ and~ ~ }\omega_{L}\rightarrow\omega_{s'}+\omega_{s''} . 
\end{eqnarray}
Here $\omega_L$ corresponds to the frequency of the observed $B_1$ mode, and $\omega_{s'}$ and $\omega_{s''}$ to the frequencies of the other two modes involved in the process.

The inverse lifetime ($\tau^{-1}$), i.e. the relaxation rate, under this SMRT approximation is given by \cite{Srivastava-Book(90)}
\begin{eqnarray}
\begin{split}
\tau_{3ph}^{-1} \propto \sum_{q's'q''s''} \big|A_{qq'q''}^{ss's''}\big|^2 \mbox{$\frac{qq'q''}{v_sv_{s'}v_{s''}}$}\delta_{q+q'+q'',G}\\
\left[\mbox{$\frac{\bar{n}_{s'}(\bar{n}_{s''}+1)}{\bar{n}_{s}+1}$}\delta(\omega_{s}+\omega_{s'}-\omega_{s''}) +\right.\\ \left.\mbox{$\frac{1}{2}$} \mbox{$\frac{\bar{n}_{s'}\bar{n}_{s''}}{\bar{n}_{s}}$}\delta(\omega_{s}-\omega_{s'}-\omega_{s''})\right].
\end{split}
\end{eqnarray}
Here $\bar{n}_{s}$ are the Bose-Einstein phonon distribution functions, $v_s$ and $q$ are the speed of sound and wavevector of the phonons, and $A$ represents the three-phonon scattering strength of the anharmonic perturbation. In Fig.\ref{figura:4} we plot with the red dashed line the result obtained within this model. Details of this calculation can be followed in the provided Supplemental Material \cite{SOM}. The remaining multiplying constant is fitted to best match the data for thicker samples, i.e. lower frequencies.  This approach yields a main dependence with $f$ in the region of interest $\tau^{-1}_{3ph}= c_1\,f$  (for this case $c_1\simeq 0.0172$), and well reproduces the experimental observation for $f<0.1$\,THz. Higher order polynomial terms are more than 10 orders of magnitude smaller than $c_1$ \cite{SOM}.

When lowering the dimensionality, i.e. restricting the acoustic propagation in one of the directions, surface effects become more important and change the dominant acoustic decay mechanism. As suggested by Balandin and co-workers \cite{Ghosh-NatureMaterials9-555(10)}, for the thinner samples, a mechanism involving the samples \emph{boundary} is expected to emerge. Following this proposal, in order to describe the observed $f^{-3}$ dependence, we model the decay using an approach that accounts for the effects of surface \textit{asperity} and the associated imperfect reflection of the vibrational acoustic modes. This simple model, proposed by Ziman back in the 60s \cite{Ziman-Book(60), Cuffe-PRL110-9(13)} considers a mean free path (MFP) $\Lambda$ of the travelling wave, determined by boundary scattering, which limits the phonon lifetime $\tau_{b}=\Lambda/v_{s}$. Within this approach $\Lambda=\frac{1+p}{1-p}\,\Lambda_o$, where $p$ represents the mean acoustic surface specularity dependent on the phonon frequency, and $\Lambda_o$ corresponds to the phonons MFP for a lossless reflecting surfaces. In our case, given the fact that the samples are thin, $\Lambda_o$ is determined by the systems characteristic dimensions, i.e. the flakes nominal thickness $\frac{\lambda_{ac}}{2}=N\,d_o$ .

Assuming small variations of the flakes thickness, and the associated surface \emph{asperity} $\eta$, defined as the root-mean-square deviation of these variations, the frequency-dependent specularity takes the form $p(f)=\exp[-16\pi^2\eta^2/\lambda_{ac}^2]$ \cite{Ziman-Book(60), Cuffe-PRL110-9(13)}. Consequently, the contribution to the lifetime due to the boundary scattering takes the form \cite{Ziman-Book(60), Cuffe-PRL110-9(13)}
\begin{eqnarray}\label{eqn-lifetime boundary}
\tau_{b} = \mbox{$\frac{\lambda_{ac}}{2v_s}$}\coth\left(\mbox{$\frac{8\pi^2\eta^2}{\lambda_{ac}^2}$}\right).
\end{eqnarray}
This expression, within a continuous elastic approximation, turns out to be proportional to $f^{-3}$ ($\lambda_{ac}^{3}$) \cite{SOM}. The inset in Fig.\ref{figura:4} displays the individual calculated values of asperity as function of $N$ obtained by using the experimental values of $\tau$ for the thinnest samples (blue squares), the acoustic speed of sound from eqn.\eqref{eqn-speed of sound linear chain model}, and eqn.\eqref{eqn-lifetime boundary}. The average asperity $\bar{\eta}\sim 2.6$\AA, is indicated by the horizontal grey line. The blue dashed curve in Fig.\ref{figura:4} shows the result using the above expression \eqref{eqn-lifetime boundary}, and the derived mean value for $\bar{\eta}$. Notice that this value for asperity represents a 20\% of a bi-layers thickness. Since the area determined by the used spot-size is small, the observed area is uniform (see Fig.\ref{figura:1}a), and the optical contrast for low $N$ is very sensitive and clearly allows to distinguish differences of one \emph{single} layer, this value for $\bar{\eta}$ might be quite reasonable. Similar phenomenology has been observed in other few-layer systems such as graphene \cite{Ghosh-NatureMaterials9-555(10)}, and can be well attributed to slight changes in the inter-atomic bonding of surface atoms, dislocations, surface wrinkling and strain, etc.\cite{Shafqat-AIPAdvances7-105306(17)}. However, further investigations need to be performed to distinguish is the origin of this asperity in of extrinsic or intrinsic nature.

For completeness, the lifetime accounting for the full acoustic frequency range can be obtained by combining both contributions, using the Matthiesen's rule $\tau^{-1}=\tau^{-1}_{3ph}+\tau^{-1}_b$ \cite{Srivastava-Book(90)}. The result for the combined lifetime is shown in Fig.\ref{figura:4} with the full grey line. As can be observed, $\tau$ very well describes the evolution of the acoustic modes lifetime in the whole frequency range, naturally, reproducing the corresponding high/low, and also the intermediate frequency region.\\

In conclusion, we have analysed using ultrafast optical spectroscopy the dynamics of longitudinal acoustic vibrational modes in high quality MoSe$_2$ exfoliated flakes, for varying thicknesses spanning from bulk like samples to few-layer systems, down to a MoSe$_2$-bilayer. The measured frequencies of the modes vary between 4.2\,GHz up to 1\,THz. By modelling the complete optical process of acoustic coherent impulsive generation and detection, we are able to precisely obtain the acoustic lifetime of the observed modes as a function of the flakes layer number, i.e. the modes frequency. A clear and strong change in the lifetime dependence with frequency is evidenced, which is associated to a dimensional crossover from a rather 3D to a 2D system. Two phenomenological models help to understand the dominating phonon scattering processes involved in each of the frequency regions, where for thicker samples the anharmonic decay via tree-phonon scattering dominates, while for thinner samples, where the acoustic mean free path becomes of the order of the flakes thickness, the dominant decay process is driven by surface boundary scattering. A combination of both models gives a quantitative description for the full span of 2D to 3D membranes.\\

Given the intimate relation of the thermal conductivity with the phonon relaxation times \cite{Maris-PhysicalAcoustics-Book(71), Ziman-Book(60)}, these results are important to understand several properties such as thermal conductivity in the stacking direction in these transition metal dichalcogenides, which are of interest for several applications. In addition, free-standing single or few layered 2D-materials, in particular 2D-TMDCs, constitute unique efficient non-linear optomechanical systems \cite{Morell-NanoLetters16-5102(16), Will-NanoLetters17-5950(17)}, and are used as high quality-factor resonators within the megahertz frequency range \cite{Morell-NanoLetters16-5102(16)}, and as tunable ultra-low mass photonic mirrors with strong and fast optical responses \cite{Back-PRL120-037401(18)}. Mechanical modes with frequencies approaching the terahertz, as those observed in this work, are optically achieved and modulate the interlayer distances. The strong excitonic resonant effects and the  large optomechanical coupling in these materials, associated to the ultra-fast strain modulation generated with relatively low-light excitation densities \cite{Mannebach-NanoLetters17-7761(17)}, together with the possibility of combining other 2D materials with different and complementary physical properties at the nanoscale via van der Waals heterostructuring \cite{Will-NanoLetters17-5950(17), Lui-PRB91-165403(15), Huang-NatureNanotechnology12-1148(17)}, opens interesting paths to establish promising opportunities for the design of devices for cavity nano-optomechanical applications \cite{Mannebach-NanoLetters17-7761(17), Gao-arXiv-1712.09245v1-26Dev2017, Weber-NatureCommunications7-12496(2016)}, eventually exploring cavity-less optomechanics \cite{Okamoto-NatureCommunications6-8478(15)}, working in the sub-terahertz regime.

\subsection*{Acknowledgements}
This work is partially supported by the Ministry of Science and Technology (Argentina) through ANPCyT grants No. PICT2015-1063. Correspondence should be addressed to A.E.B.


\newpage


%


\onecolumngrid
\newpage
\setcounter{figure}{0}  
\renewcommand{\figurename}{FIG. \!}
\renewcommand{\thefigure}{S\arabic{figure}}


\begin{center}
\underline{\huge{\textbf{Additional Information}}}
\end{center}
\vspace{0.5cm}
This supplementary information extends some points discussed in the main text, and presents additional data that could be of interest for some readers. 
Section \ref{1} discusses the sample, addressing how the ``free-standing'' (unsupported) samples are distinguished. 
The processing and analysis of the measured $\Delta R/R$ transients are explained in section \ref{3}.
The reason for the chosen pump-probe fluence used in the experiment is explained in section \ref{2}. 
Section \ref{sec-LCM} provides a brief description of the linear chain model used to describe the vibrational modes in combination to additional Raman experiments in few-layer MoSe$_2$ to derive the out-of-plane acoustic sound velocities.
The particular case of the thicker bulk-like samples, their analysis and simulations are described in section \ref{4}. 
In section \ref{5} we address the model used to simulate and fit the anharmonic scattering decay rate of acoustic phonons in bulk MoSe$_2$ through an adapted anisotropic Debye approximation, and in section \ref{sup} some considerations related to the surface roughness phonon decay mechanism, of main importance in the thinner samples, are analyzed. 
Finally, a brief analysis of the quality factor of the $B_1$ modes, assumed as a simple mechanical oscillator, is presented in section \ref{6}.

\section{Sample addressing}\label{1}

\begin{figure}[bbb]
\includegraphics*[keepaspectratio=true, clip=true, angle=0, width=.6\columnwidth, trim={0mm, 20mm, 0mm, 25mm}]{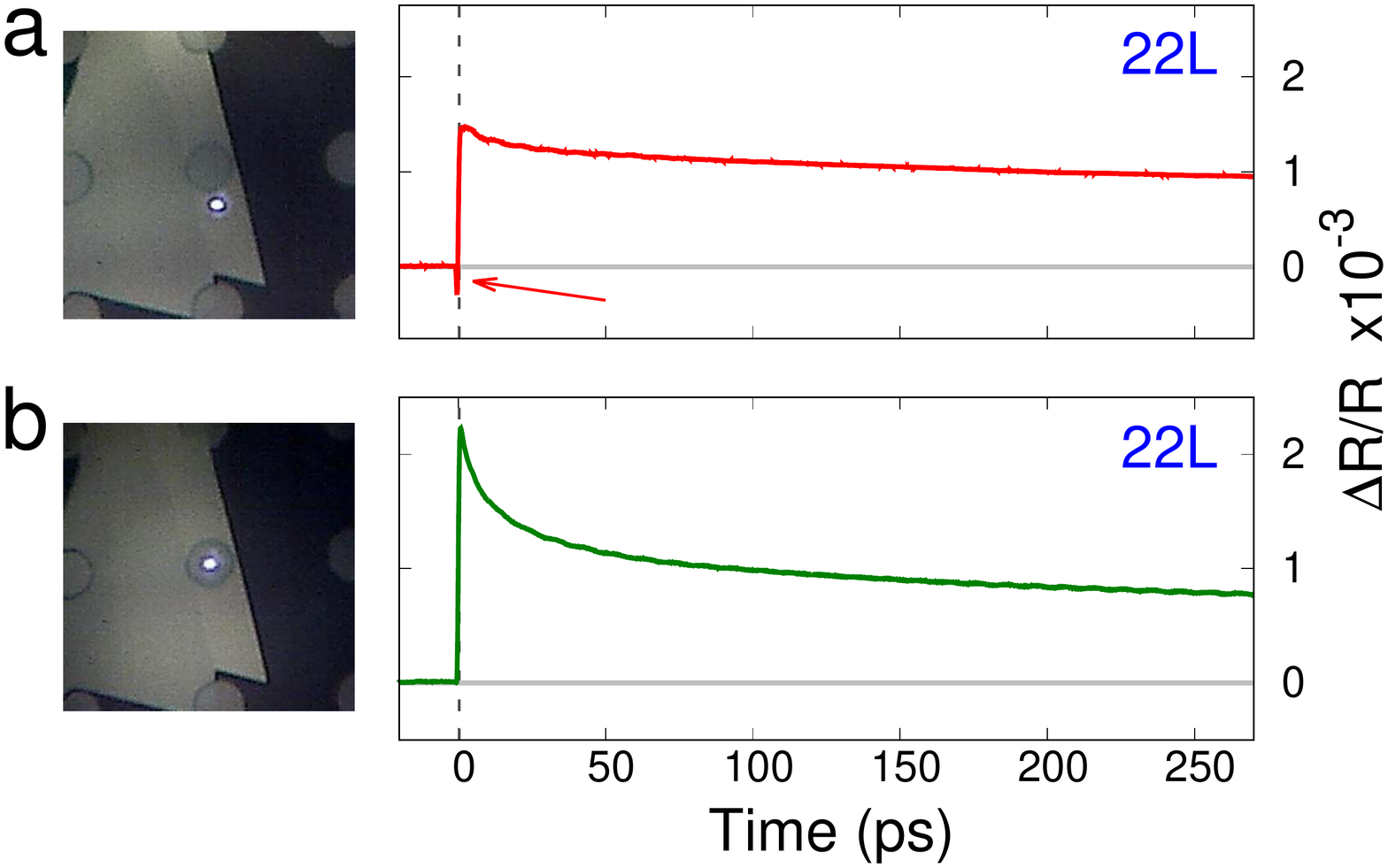}
\caption{Typical ``as measured'' transient reflectivities $\Delta R/R$ for a supported sample (\textbf{a}) and a suspended sample (\textbf{b}). The optical images left to the transients, taken through the 100$\times$ objective, show the precise addressing of the laser spot on top of the sample for the respective cases.}
\label{figura:SM1}
\end{figure}

The exfoliated MoSe$_2$ flakes are scattered on a 90\,nm SiO$_2$/Si substrate wafer, which has been previously patterned with regular circular holes of $\sim$6\,$\mu$m in diameter. As a result, some flakes were randomly deposited on top of the holes, and were therefore only supported by the holes' borders, i.e. were in a ``free-standing'' condition. 

To ensure the correct addressing of the spot on top of the samples, our experimental system allows to acquire simultaneously the white light image and the time-resolved measurement. Figure \ref{figura:SM1} on the left, shows the optical microscope images for a 22L MoSe$_2$ flake through the 100$\times$ objective, where the laser spot is clearly observed next to a patterned hole (Fig. \ref{figura:SM1}a, left), and centred on top of this hole (Fig. \ref{figura:SM1}b, left). The corresponding measured transient reflectivity $\Delta R/R$, for each condition, is shown on the respective right panels. The origin of the observed transient signal is mainly due to the changes in the optical constants resulting from the impulsively modified electronic states within the samples at $t = 0$, i.e. at the arrival of the pump pulse. 

Both transient signals are relatively strong, but evidently different for each of the situations. On top of the hole, the onset (at $t=0$) changes abruptly from its unperturbed position towards positive values (Fig. \ref{figura:SM1}b, right) and relaxes back towards its equilibrium. Besides the hole, within the first few picoseconds, the transient reflectivity has an initial impulsive deflection towards \emph{negative} values (see the arrow in Fig. \ref{figura:SM1}a, right), and rapidly changing to positive values. This initial negative impulsive deflection, the slightly lower initial positive value reached before the signal's quasi-exponential decay towards equilibrium, and the smaller exponential decay are a systematic behaviour displayed by the transients obtained for the supported flake, and is a fingerprint -in addition to the optical image- that enables unambiguously to distinguish in which situation we are, i.e. if the spot is on a supported or suspended (free-standing) position in the flake.

\section{Signal processing}\label{3}

\begin{figure}[bbb]
\includegraphics*[keepaspectratio=true, clip=true, angle=0, width=.6\columnwidth, trim={0mm, 15mm, 0mm, 25mm}]{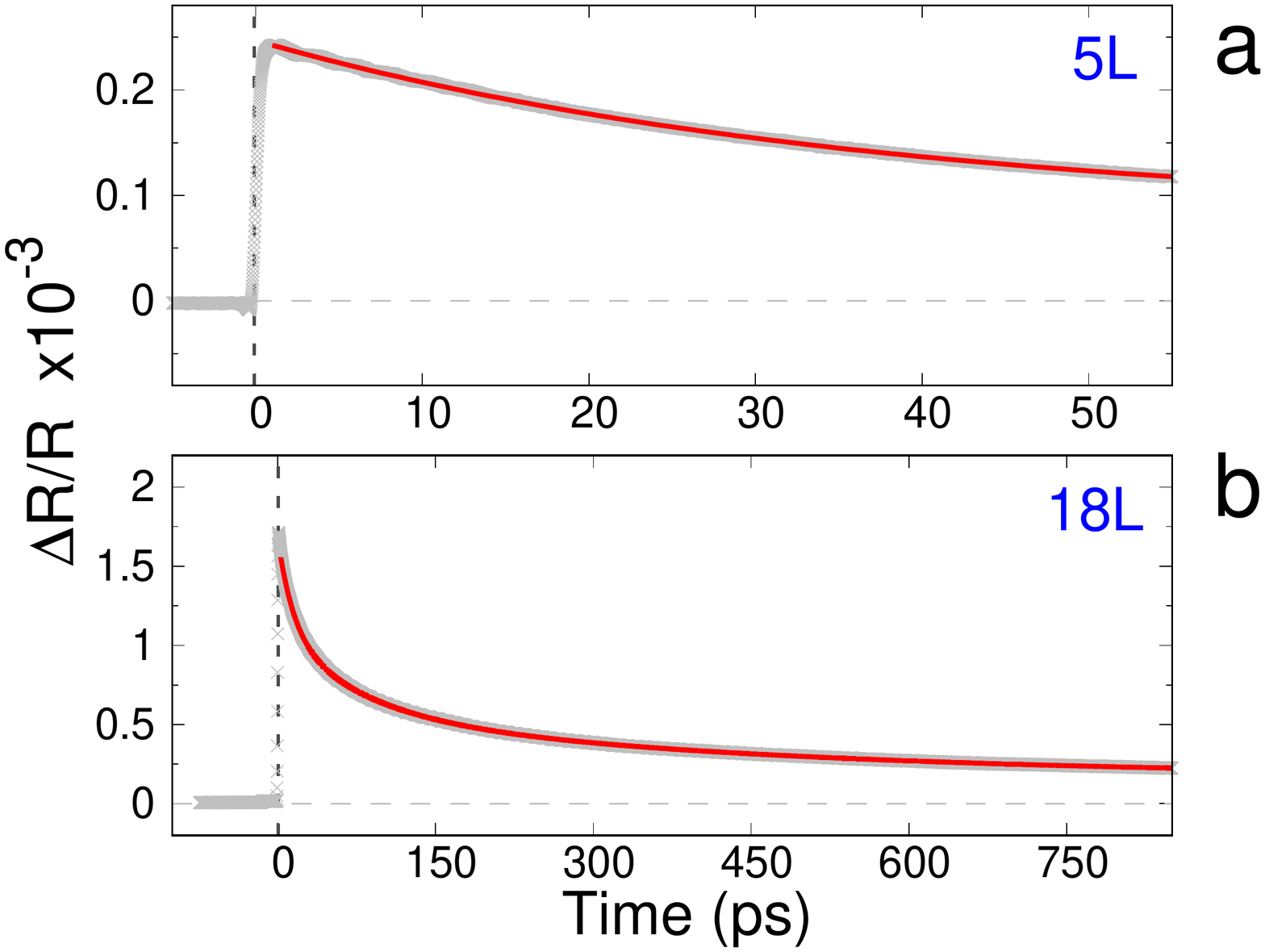}
\caption{As measured $\Delta R/R$ transient for samples with 5L MoSe$_2$ (\textbf{a}) and 18L MoSe$_2$ (\textbf{b}). The grey symbols correspond to the measurement and the red curves are the multi-exponential fitting. For the 18L sample the sum of three decaying exponentials was needed to fit the experiment, while for the 5L sample only one decaying exponential was enough to well describe the curve.}
\label{figura:SM3}
\end{figure}

Figure \ref{figura:SM3} shows the untreated (``as measured'') $\Delta R/R$ transient for 5L and 18L MoSe$_2$, panels a) and b) respectively. The contribution to the transient reflectivity due to the phonon modes in the sample needs to be extracted by subtracting a fitted multi-exponential function, where the exponential decay constants are related with electronic processes, such as the decay of excitons and intervalley scattering, or exciton dephasing \cite{Jeong-ACSNano10-5560(16)}. The red curves in the figure \ref{figura:SM3} are the corresponding multi-exponential fittings and show a good agreement with the experiment. It is important to point out that the measured temporal window for each experiment was modified depending on the phonon dynamics, extending the window when the phonon lifetime was longer, reducing the relative noise of the measurement. To perform the fitting, we used the sum of up to three decaying exponential functions to have the best possible description, and consequently the cleanest contribution of the phonon mode oscillations. For the sample with 5L a single exponential is enough to well fit the decay ($\tau_1\simeq 37$\,ps). For the sample with 18L, three exponential functions were needed to describe the curve. For the later case, the best agreement was found using: $\sum_{m=1}^{m=3} A_m \exp(-t/\tau_m)$, with $\frac{A_2}{A_1}=0.890$, $\frac{A_3}{A_1}=0.711$, $\tau_1=14.8$\,ps, $\tau_2=75.3$\,ps, and $\tau_3=432$\,ps.\\

\begin{figure}[tt!!!]
\includegraphics*[keepaspectratio=true, clip=true, angle=0, width=.6\columnwidth, trim={0mm, 0mm, 0mm, 0mm}]{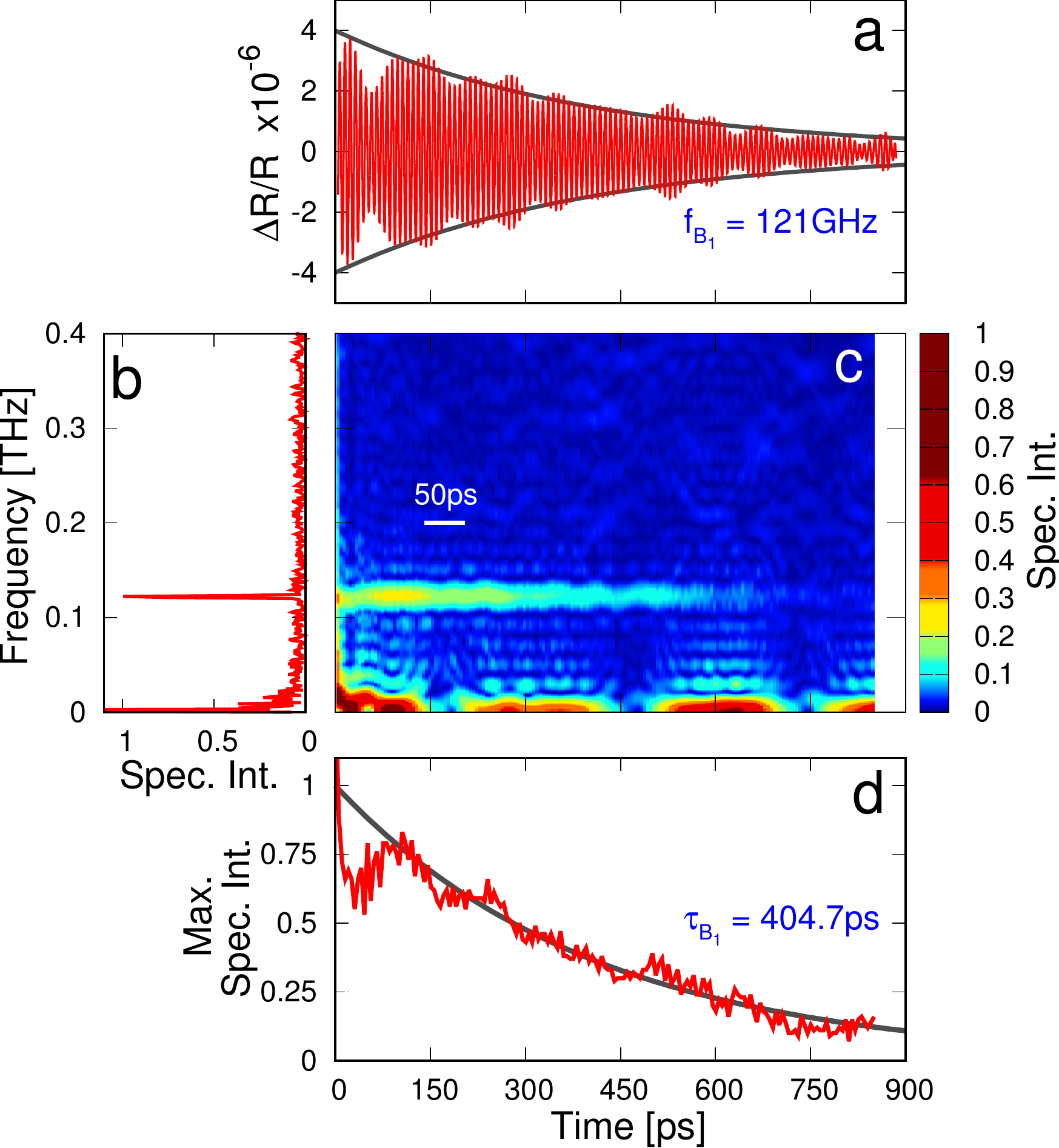}
\caption{Signal processing for 18L MoSe$_2$. \textbf{a)} filtered $\Delta R/R$ (red) and fitted exponential decay envelope (grey). \textbf{b)} numerical Fourier transform of $\Delta R/R$. \textbf{c)} density map of the wnFT of the phonon signal $\Delta R/R$ (gliding window 50\,ps). \textbf{d)} $B_1$ mode's intensity extracted from the density map of the wnFT (red) and fitted exponential decay (grey).}
\label{figura:SM4}
\end{figure}

The extracted oscillations for the samples with 5L is shown in Figure 2 of the main text, and for the case of the 18L MoSe$_2$ sample, the result and consequent treatment obtained after the subtraction process described above is shown in Fig.\ref{figura:SM4}.

The filtered phonon contribution to the transient is displayed in Fig.\ref{figura:SM4}a. It is possible to clearly observe the decrease of the acoustic oscillation's amplitude associated to the phonon damping. The exponential envelope is indicated (gray curve) and  a $\tau_{B_1}$ of 404.7\,ps is obtained.

The spectral components of this signal can be better analyzed by performing the numerical Fourier transform (nFT), which is presented in Fig.\ref{figura:SM4}b. The spectrum displays a \emph{single} peak at $\sim$121\,GHz corresponding to the $B_1$ mode of the suspended flake. To have a hint of the temporal dynamics in the spectral domain, a windowed numerical Fourier transform (wnFT) is presented in Fig.\ref{figura:SM4}c. This wnFT was performed using a gliding window of 50\,ps, as indicated in the figure. This density plot shows the behavior of the intensity of each spectral components and its evolution in time. The intensity of the $B_1$ mode at $\sim$121\,GHz decreases exponentially, vanishing below the noise level. Figure \ref{figura:SM4}d displays the $B_1$ mode intensity extracted from the density map of the wnFT together with the same exponential envelope presented in panel a) and yielding the same acoustic exponential decay time. The low frequency noise appearing in Fig.\ref{figura:SM4}c for frequencies $\leq$ 25\,GHz, are probably an artifact that remains from subtracting the electronic contribution. 

\section{Pump laser fluence dependence}\label{2}

\begin{figure}[t!!]
\includegraphics*[keepaspectratio=true, clip=true, angle=0, width=.5\columnwidth, trim={0mm, 15mm, 0mm, 21mm}]{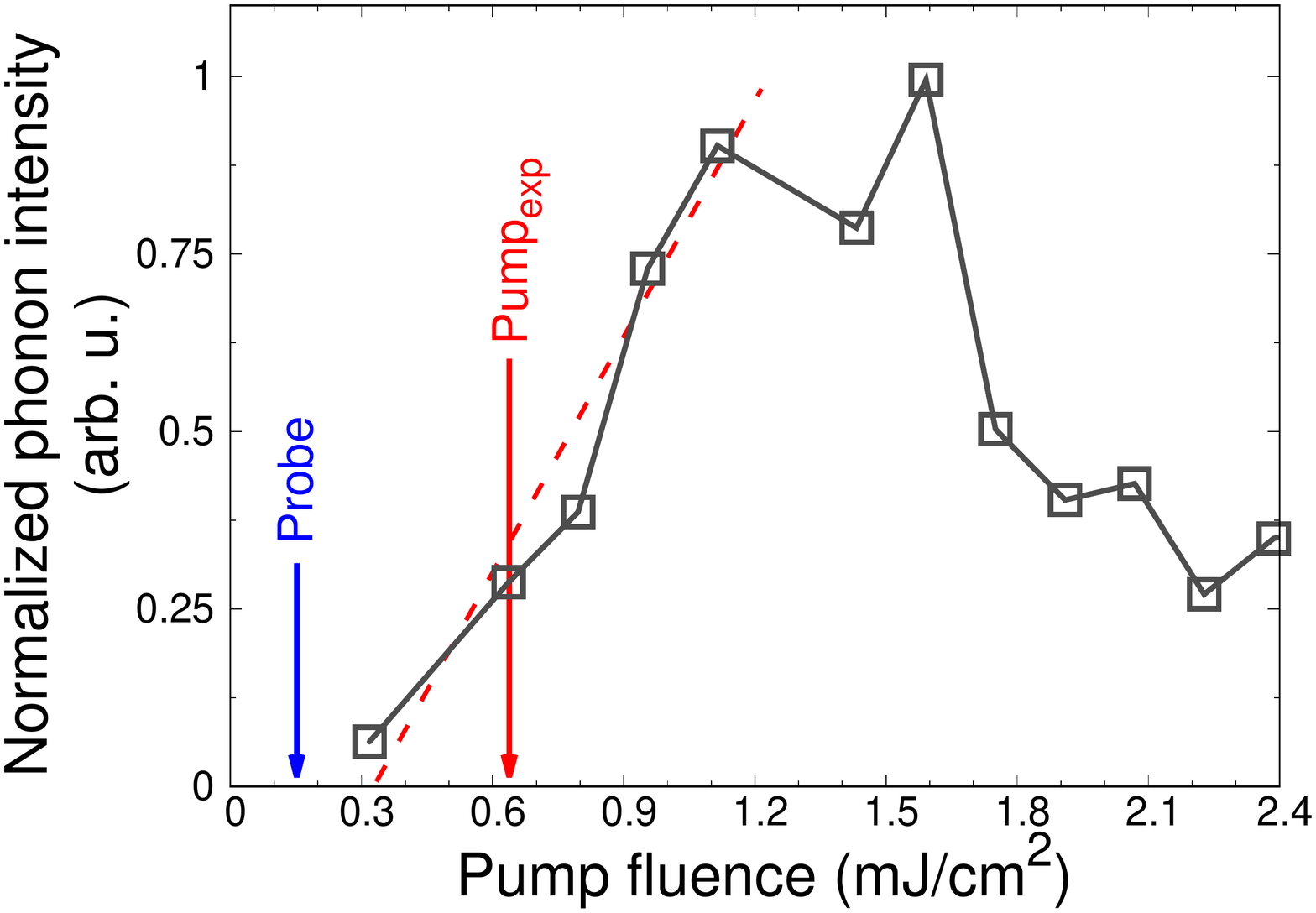}
\caption{Fluence dependence of the $\Delta R/R$ intensity of the $B_1$ mode with the pump laser. The fluence of the probe beam was set to 0.16\,mJ/cm$^2$ (blue arrow). The red dashed line shows the approximately linear dependence, and the red arrow indicates the pump laser fluence used in the phonon lifetime measurements.}
\label{figura:SM2}
\end{figure}

All the transient reflectivity pump-probe measurements presented were performed with a 100$\times$ objective of NA=0.95, that was used to focus the laser down to a spot of $\sim$1\,$\mu$m diameter. Figure \ref{figura:SM2} shows the pump laser fluence dependence of the spectral intensity of the fundamental mode, for a 22 layer MoSe$_2$. The blue vertical arrow indicates the fluence of the probe laser used in all the measurements, which was set to 0.16\,mJ/cm$^2$. As observed, the phonon intensity grows rather linearly with the generating incident power until its maximum at about $\sim$1.20\,mJ/cm$^2$. When further increasing the pump power, the spectral intensity saturates and decreases afterwards until $\sim$3\,mJ/cm$^2$, where the sample breaks. The red dashed line indicates the approximately linear  within the initial range. 
The pump fluence used in the experiments for the phonon lifetime determination, is marked by the red arrow, and was set to 0.64\,mJ/cm$^2$.
It was chosen to be significantly larger than the probe power, but far enough from the saturation. The pump-probe fluence ratio (4:1) is coincidently similar to those values used in other reported works for experiments that are alike \cite{Jeong-ACSNano10-5560(16), Ge-ScientificReports4-5722(14)}. 

\section{Linear chain model for interlayer modes} \label{sec-LCM}

\begin{figure}[bbb]
\includegraphics*[keepaspectratio=true, clip=true, angle=0, width=.6\columnwidth, trim={0mm, 0mm, 0mm, 0mm}]{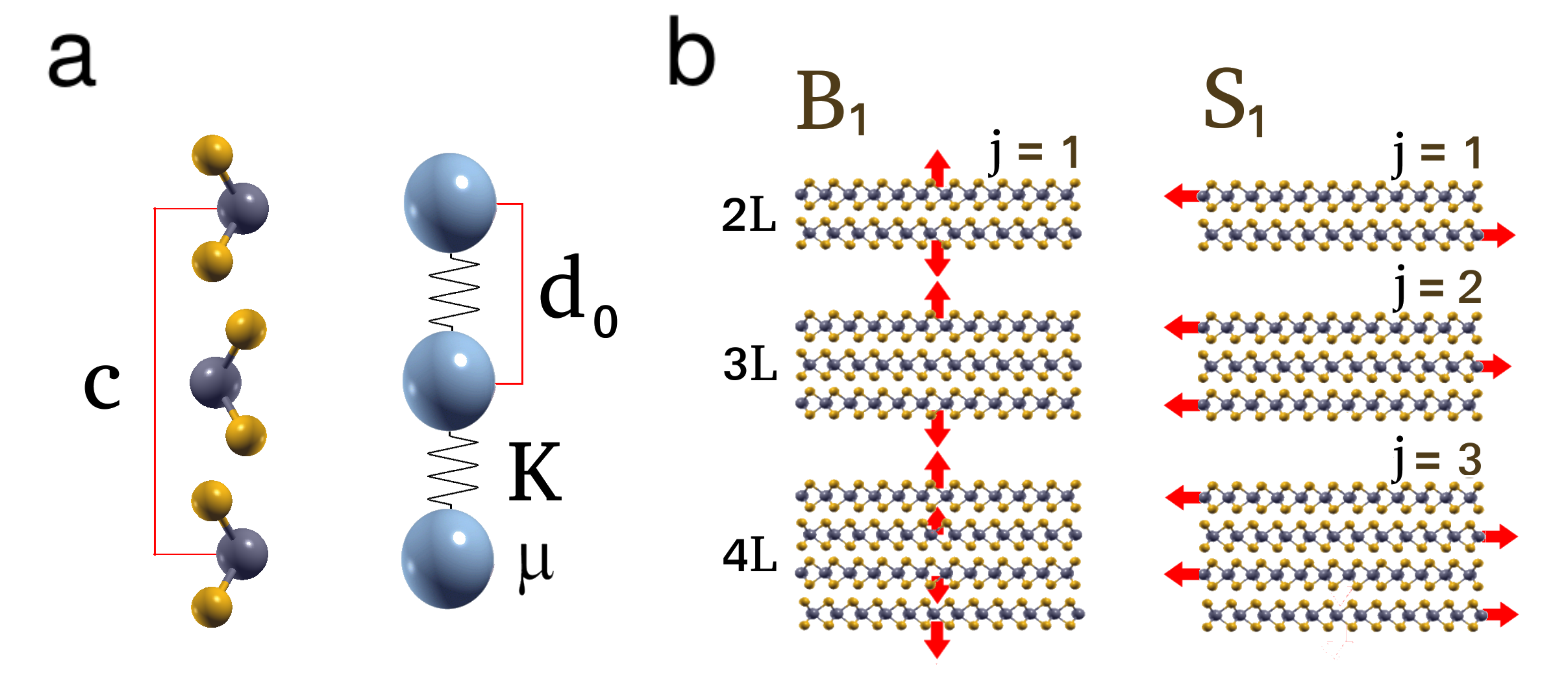}
\caption{\textbf{a)} linear chain model, each layer is replaced by an equivalent effective mass $\mu$ per unit area and a nearest-neighbouring interlayer force constant $K$ per unit area. \textbf{b)} schematic of interlayer $S_1$ and $B_1$ modes for $N=2,3$ and 4, the arrows indicate the vibration direction of each rigid layer.}
\label{figura:SM10b}
\end{figure}

The 2$H$-MoSe$_2$ is a two dimensional material in which the atoms within each layer are connected by covalent bonds while the bulk crystal is formed by the stacking of these layers via van der Waals interactions. The interlayer shear ($S$) and the interlayer breathing ($B$) modes are characterized by the relative motion of the different layers but leaving the internal structure within each individual layer intact \cite{Ji-PhysicaE80-130(16), Liang-ACSNano11-11777(17)}. The phononic properties of a multilayer MoSe$_2$ depend critically on the number of layers $N$ and thereby, this make Raman scattering a useful technique to characterize these two-dimensional materials \cite{Liang-ACSNano11-11777(17), Soubelet-PRB93-155407(16), Lu-NanoResearch9-3559(16), MolinaSanchez-SurfSciRep70-554(15)}.

Under a simple linear chain model to describe the shear and breathing modes, the atomic details within each layer are not necessary and are replaced by parameters that characterize the interlayer van der Waals interactions. As it is presented in figure \ref{figura:SM10b}a, each layer of the material is replaced by an equivalent effective mass $\mu$ per unit area and a nearest-neighbouring interlayer force constant $K$ per unit area. This approximation implies the substitution of the original layered crystal for a chain of effective masses. Since the acoustic modes do not imply the relative motions of atoms within each layer, each unit cell of the material could be analyzed as having two effective masses ($n_0 = 2$). It is clear that this approximation cannot describe the optical phonons of the system, that imply the relative motion of atoms within each layer.

Solving the elastic equation of motion for the linear chain model and accounting free surface (stress-free) boundary conditions, results the well known dispersion relation for the frequency of the modes as function of the number of layers: \cite{Ji-PhysicaE80-130(16), Liang-ACSNano11-11777(17), Liang-Nanoscale9-15340(17)}
\begin{eqnarray}\label{LCM}
f_{B(S)_{N,j}} = f_{B(S)_0}\sin\left(\frac{k_{N,j}d_0}{2}\right),
\end{eqnarray}
for the $B$($S$) mode. The interlayer distance is $d_o\simeq c/2$, where $c=12.918$\,\AA\ is the bulk lattice parameter in the stacking direction \cite{Roy-ACS-AppMat8-11(16), Coehoorn-PRB35-12(87)}, $k_{N,j}=\frac{2\pi}{\lambda_{ac}}$ is the acoustic wavevector for the associated wavelength $\lambda_{ac}=\frac{2 N d_0}{j}$ and $j$ the phonon branch index. $f_{B(S)_0}$ is related to the interlayer force constant per unit area $K_{\perp}$($K_{\parallel}$) and $\mu$ as
\begin{eqnarray} \label{eqn - f_o}
f_{B(S)_0}=\sqrt{\mbox{$\frac{K_{\perp(\parallel)}}{\pi^2 \mu}$}}.
\end{eqnarray}\\

\begin{figure}
\includegraphics*[keepaspectratio=true, clip=true, angle=0, width=.7\columnwidth, trim={0mm, 0mm, 0mm, 0mm}]{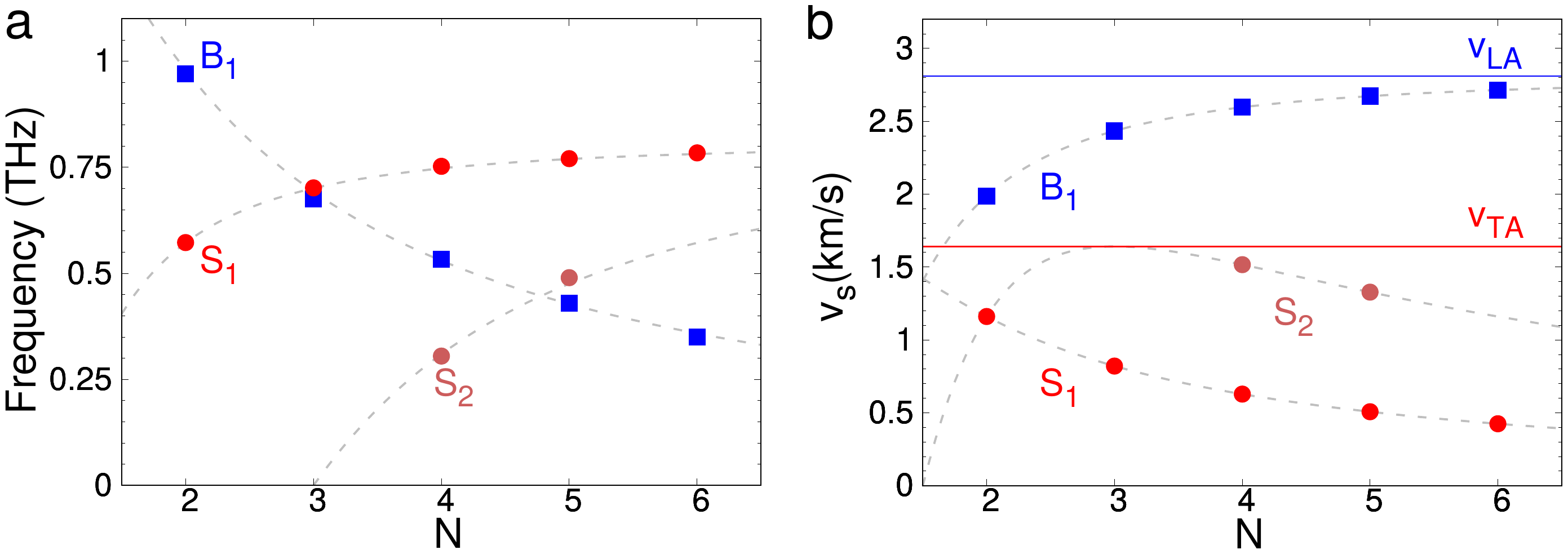}
\caption{\textbf{a)} $B$ and $S$ mode energies for Raman experiments performed for the thinner samples. \textbf{b)} calculated velocities for the corresponding $B$ and $S$ modes identified by Raman. The interpolation of the fitted values is shown with dashed grey lines and the blue(red) line is the bulk speed of sound for longitudinal(transversal) modes.}
\label{figura:SM10}
\end{figure}

The identification of the $B$ and $S$ modes in Raman scattering requires the measurements of polarized and cross-polarized Raman spectra. According to the Raman tensors, while the $B$ mode could be observed only under parallel polarization measurements, the $S$ mode is observed in both configurations \cite{Liang-ACSNano11-11777(17)}. Figure \ref{figura:SM10}a shows the $B$ and $S$ mode energies for Raman experiments performed in the thinner samples ($N = 2,3,4,5$ and $6$). The observed $S_1$ and $S_2$ modes belong to branches whose frequency increase with increasing $N$ and correspond to the phonon branches $j=N-1$ and $j=N-3$ respectively. The $B_1$ mode instead belong to branches whose frequency decreases with increasing $N$ and correspond to $j=1$. This observations are in accordance with experiments performed in others TMDCs \cite{Liang-ACSNano11-11777(17), Liang-Nanoscale9-15340(17)}. Figure \ref{figura:SM10b}b presents the schematics of interlayer $S_1$ and $B_1$ modes, the arrows indicate the vibration direction of each rigid layer.

The results presented in Fig. \ref{figura:SM10}a are fitted with the expression \eqref{LCM}, where the only free parameter corresponds to $f_{B_0}$ for the $B$ mode and $f_{S_0}$ for the $S$ mode. The interpolation of this fitted curve is shown with dashed grey lines. The fitted values are $f_{B_0}=(1.36\pm 0.03)$\,THz and $f_{S_0} = (0.81\pm 0.03)$\,THz. An estimation for the in-plain effective mass for each layer, accounting the atomic masses and the MoSe$_2$ in-plane unit cell \cite{Roy-ACS-AppMat8-11(16), Coehoorn-PRB35-12(87)}, results $\mu\simeq 4.41\times 10^{-6}$\,kg/m$^2$. From the above fit, we can estimate using eqn.\eqref{eqn - f_o} the effective interlayer elastic force constants $K_{\perp}=8.42\times 10^{18}$\,N/m$^3$ and $K_{\parallel}=2.85\times 10^{18}$\,N/m$^3$. These values are of the order of those obtained for similar TMDCs 2D-systems \cite{Froehlicher-NanoLetters15-6481(2015), Zhao-NanoLetters13-1007(13)}.

The acoustic propagation group velocity $v_{B(S)s}$ could be calculated from \eqref{LCM} applying the derivative with respect to the wavevector as $v_{B(S)s} = \frac{d\omega}{dk}$, and results
\begin{eqnarray}\label{V-LCM}
v_{B(S)s}(N,j)= \pi f_{B(S)_0}d_0\cos\big(\mbox{$\frac{k_{N,j}\,d_o}{2}$}\big)\ .
\end{eqnarray}
Figure \ref{figura:SM10}b shows the calculated velocities for the corresponding $B$ and $S$ modes identified by Raman. The dashed grey curves show the interpolation of eqn.(\ref{V-LCM}). As the $B$ and the $S$ modes belong to different kind of branches, the regarding group velocity have different behaviour. The speed of sound along the stacking direction tends to the bulk speed of sound for the branch of index $j=1$. For this reason, the $v_{B}$ grows as function of $N$ asymptotically to the value of the longitudinal acoustic velocity along the stacking direction ($v_{LA}$), while $v_{S}$ decreases tending to zero when increasing $N$. The ``bulk'' longitudinal(transversal) acoustic velocity along the stacking direction $v_{LA}$($v_{TA}$) could be obtained as a limit of $v_B$($v_S$) [eqn.(\ref{V-LCM})] for $j=1$ and $N\rightarrow\infty$. The estimated ``bulk'' speed of sound are $v_{TA} = 1630$\,m/s and $v_{LA} = 2820$\,m/s ($v_{ac}$ in the main text). The latter is in accordance with the pump-probe measurements (see main text, Fig.\,3).

\section{Pump-probe experiments in thick samples}\label{4}

By ``thick samples'' we mean thicknesses ($d$) of the flakes that are of the order or larger than the optical penetration depth
\begin{eqnarray}\label{alpha}
\delta_p = \frac{\lambda_{op}}{4\pi \Im m (\widetilde{n})},
\end{eqnarray}
where $\lambda_{op}=805$\,nm is the central wavelength of the laser and $\widetilde{n}$ the complex refractive index taken from Ref.[\onlinecite{Li-PRB90-205422(14)}]. It is in these conditions that the model used for the simulations displays mayor advantages over an empirical function of the form $U(t) = A\exp(-t/\tau)\sin(2\pi ft+\phi)$, as it is usually used to derive the amplitude, frequency, and phase of the modes. The pump-pulse, roughly following its absorption profile, excites longitudinal acoustic strain pulses, which are launched into the sample. Analogously, the detection is thus sensitive to the presence of this propagating strain pulse within the volume given by the optical penetration depth of the probe laser \cite{Thomsen-PRL53-989(84), Thomsen-PRB34-4129(86)}. 

Such propagating strain pulse, needs to be described as a certain superposition of the vibrational eigemodes of the flake (see Section \ref{sec-LCM}). The important consequence is that the $\Delta R/R$ resulting from a combination of several modes cannot be described by the simple empirical $U(t)$ function, and mandatory needs to include the effect of the higher frequency modes that are excited within the flake. For consistency, we have used the \emph{same} more complex model \cite{PascualWinter-PRB85-235443(12)} that fully calculates $\Delta R/R$ to analyse all the measurements, and leaving the sample thickness (number $N$ of layers) and the phonon lifetime ($\tau_{B_1}$) as fitting parameters.

In what remains of this section we will show two cases where it becomes evident that the simple empirical $U(t)$ function fails to describe the observations: The first case corresponds to a MoSe$_2$ flake of $N=57$L, leading to a thickness of $d=36.8$\,nm ($\delta_p \sim 100$\,nm). And the second, corresponds to the thickest MoSe$_2$ sample found, $N\simeq519$L ($d=335$\,nm $\gg$ $\delta_p$), and basically behaves as ``bulk'' MoSe$_2$.\\

\subsection{57L MoSe$_2$}
\begin{figure}
\includegraphics*[keepaspectratio=true, clip=true, angle=0, width=.6\columnwidth, trim={0mm, 20mm, 0mm, 16mm}]{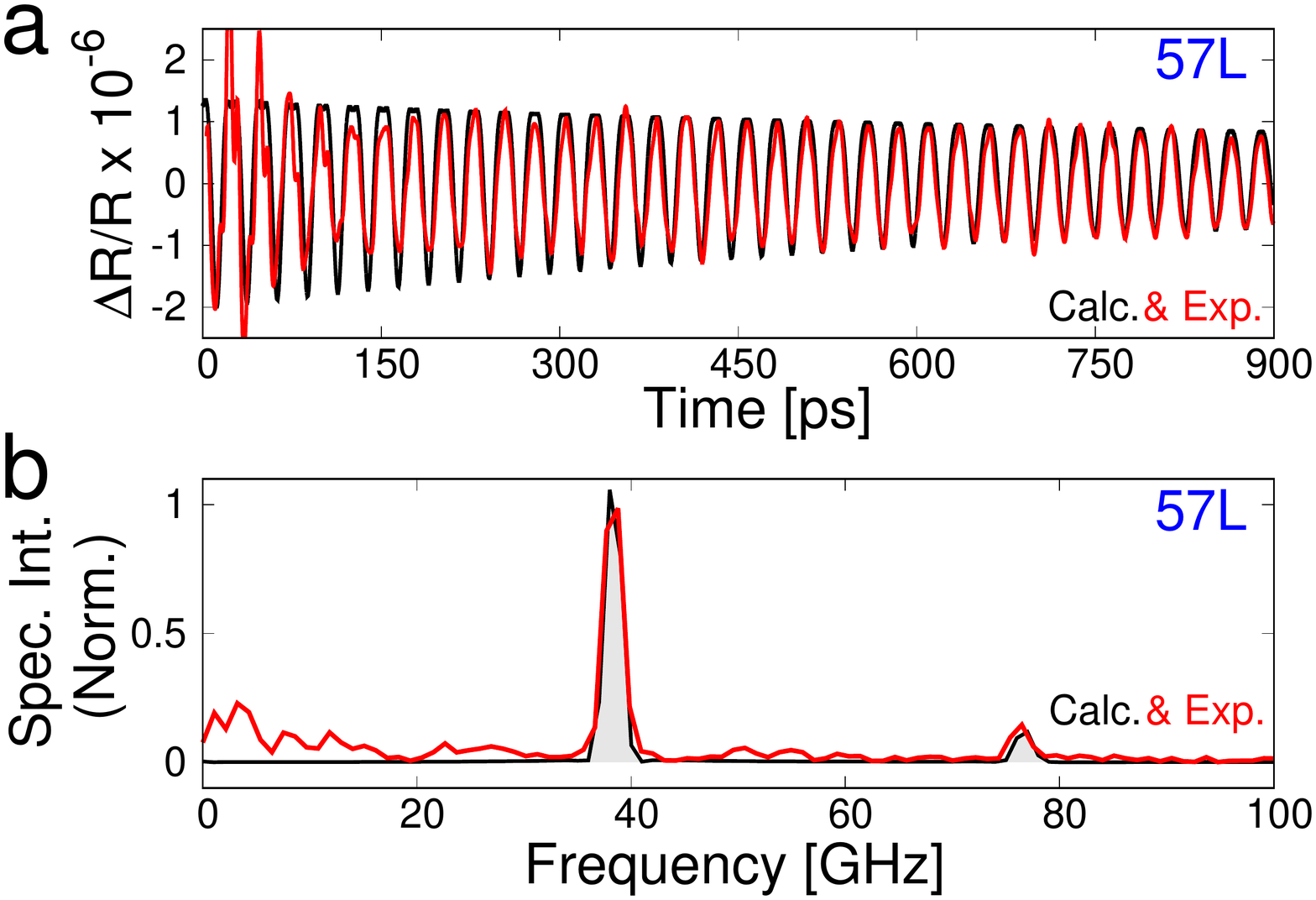}
\caption{\textbf{a)} Extracted oscillations (red curves) in the time-domain showing, corresponding to a 57L-MoSe$_2$ sample. \textbf{b)} Normalized numerical Fourier transform of the previous transient. The black curves are the simulations that best fit the experiments simultaneously in both domains.}
\label{figura:SM6b}
\end{figure}
Between a thin flake and the bulk case, there are sample thicknesses that could not be catalogued in neither of these two groups. This is the case where the penetration depth is of the order of the MoSe$_2$ sample. 

Figure \ref{figura:SM6b}a shows the measurement (red) and the simulation (black) for a 57L MoSe$_2$ free-standing sample. After the first $\sim150$\,ps, where the signal has kind of an irregular behaviour, the simulation has a relatively good agreement with the experiment. The nFT of this transient is presented in figure \ref{figura:SM6b}b and the calculated curve is very well reproduced.  Two peaks dominate the spectrum. The intense peak at $\sim$38\,GHz corresponds to the fundamental confined mode, while the weaker one at $\sim$76\,GHz to its second harmonic. This situation is clearly an intermediate one, where the penetration depth is of the order of the flakes thickness. Here the asymmetry of the induced initial stress is responsible for the generation of both, odd and even modes, as observed.

As we have chosen for the simulation the phonon lifetime proportional to $1/f$, the second harmonic mode at $\sim$76\,GHz presents a faster relaxation time, that is observed in time domain for the lower times in the simulations and also in the measurement. In the spectral domain, the faster relaxation time for the higher frequencies is responsible for the intensity ratio between the peaks, that fits with the experiment. The value of $N = 57 \pm 2$ layers MoSe$_2$ together with the $B_1$ relaxation time $\tau = 1.52\pm 0.03$\,ns are critical for the fitting of this model. Note the excellent agreement of experiment and theory.
								
\subsection{``Bulk'' MoSe$_2$}
 
\begin{figure}
\includegraphics*[keepaspectratio=true, clip=true, angle=0, width=.6\columnwidth, trim={0mm, 20mm, 0mm, 16mm}]{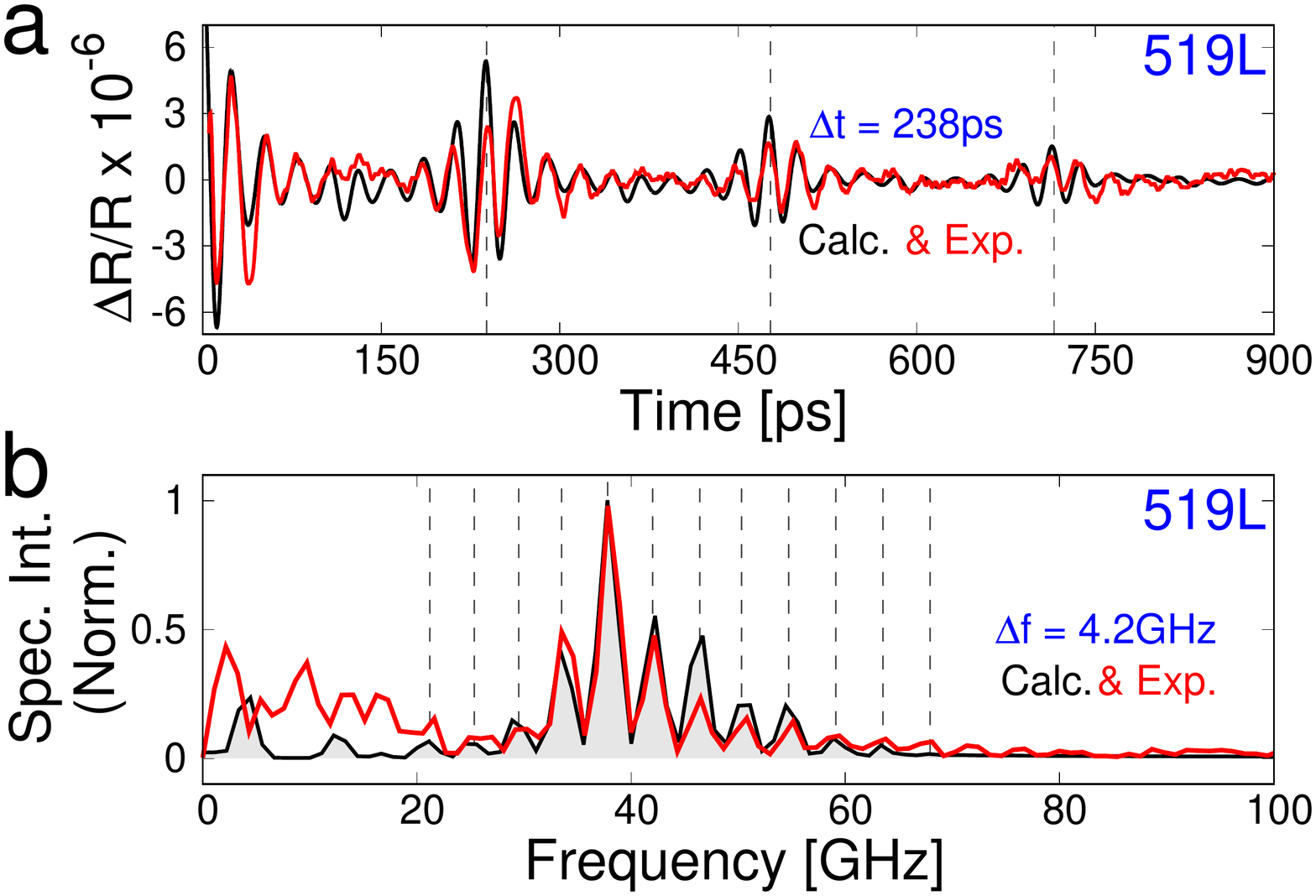}
\caption{\textbf{a)} phonon signal (red curves) in time-domain showing the extracted oscillations corresponding to a thick sample (519 layers MoSe$_2$). \textbf{b)} Numerical Fourier Transform of the corresponding transients. The black curves are the simulations that best fit the experiments simultaneously in both domains.}
\label{figura:SM6}
\end{figure}

Figure \ref{figura:SM6}a presents the extracted $\Delta R/R$ (red) in time domain for the pump-probe measurement of this sample. The signal does \emph{not} correspond to a beating but to multiple reflections of the generated acoustic pulse at the flake's back and front surfaces of the sample, i.e. the MoSe$_2$-SiO$_2$ and the air-MoSe$_2$ interfaces. The acoustic pulse is generated at the flakes surface and propagates into the sample with a speed $v_{LA}$. As the pulse penetrates the sample, since the penetration of light decreases exponentially with depth, the detection is diminished (signal $\Delta R/R$ decreases). At around 124\,ps the pulse gets reflected back at the back side of the sample and returns to the surface. By doing so, the acoustic pulse reenters the region where the probe laser is again sensitive to its detection ($\Delta R/R$ increases) \cite{Thomsen-PRL53-989(84)}. At $t\simeq$ 248\,ps the acoustic pulse is reflected at the air-MoSe$_2$ interface and is again directed into the sample, were the process is repeated. It turns out that within this regime, the vibrational modes that are addressable have the frequencies described by \cite{Thomsen-PRL53-989(84)}
\begin{eqnarray}\label{eqn:Brillouin frequency}
f_b \simeq 2\,k_{op}\,v_{LA}\sim \frac{4\pi v_{LA} \Re e(\textrm{\~n})}{\lambda_{op}},
\end{eqnarray}
where $k_{op}$ is the optical wavevector of the probe laser. In the literature, the acoustic pulse are usually called ``Brioullin'' mode and is generated with the allowed frequency modes in the sample described by eqn.(\ref{LCM}).

In the limit $N \gg 2$ the allowed modes within the sample are equi-spaced defining the \textit{free spectral range} (the acoustic mode's separation) \cite{Grossmann-PRB-88-205202(13)}
\begin{eqnarray}\label{fsr}
\Delta f = v_{LA}/(2d)\ ,
\end{eqnarray}
where $\Delta f$ matches the lower mode frequency ($B_1$). Figure \ref{figura:SM6}b presents the Fourier analysis of the signal shown on \ref{figura:SM6}a. The nFT basically shows a set of peaks with the frequency of the modes existing in the acoustic pulse and separated by $\Delta f = 4.2$\,GHz, that according to eqn.(\ref{fsr}) implies $d \sim 335$\,nm $\simeq$ 519 layers of MoSe$_2$.  

In addition, considering the pulse travelling time of $\Delta t\sim 238$\,ps within the samples, using the sound velocity obtained in section \ref{sec-LCM}, one can estimate the thickness of this particular flake being 
\begin{eqnarray}
d=\mbox{$\frac{\Delta t}{2}$}\, v_{ac}\sim 333\,\text{nm}\simeq 516~\text{layers MoSe$_2$}\ .
\end{eqnarray}
This value is reasonable and not far from the thickness obtained before. 

The simulation of $\Delta R/R$ performed for this sample is presented in figure \ref{figura:SM6} with black curves, and shows a good agreement with the experiment. The frequency modes and the multiple reflections of the acoustic pulse at the flake's surfaces of the sample are well reproduced in the temporal and spectral domain. As mentioned, the thickness of the MoSe$_2$ is a critical fitting parameter. Here we obtained $N = 519 \pm 5$ layers MoSe$_2$ and is also in accordance with the estimations above.

As is also evidenced here, the complex behaviour of the signal cannot be reproduced by the empirical function $U(t)$, and it is necessary to introduce the model for $\Delta R/R$ to correctly interpret the experimental results.

\section{Anharmonic scattering mechanism for the decay of phonons in anisotropic materials}\label{5}

The description of a crystal lattice accounting only a harmonic potential can not reproduce the phonon-phonon interaction, which provides one main mechanism responsible for the phonon decay in pure crystals. The relaxation time of a phonon due to three-phonon processes arises from the inclusion of \textit{anharmonic} terms in the lattice Hamiltonian, casting away the concept of non-interacting phonons \cite{Srivastava-Book(90), Cuffe-PRL110-9(13), Maris-PhysicalAcoustics-Book(71)}.\\

One model that captures the essence of these kind of three-phonon processes for calculating the acoustic decay times, is the single mode relaxation time (SMRT) approximation \cite{Srivastava-Book(90), Cuffe-PRL110-9(13), CallawayPR.113.1046(59)}. In particular, this approach using a Debye-type approximation proved to give reasonable results in similar systems for low frequency acoustic phonons \cite{Srivastava-Book(90), Cuffe-PRL110-9(13)}. The conventional Debye model works fine for isotropic systems, where the velocities of sound are equal in all directions. Cases such as bulk one-dimensional Van der Walls crystals, e.g. bulk MoSe$_2$ or graphite, have quite different in-plane/out-of-plane velocities of sound and are thus anisotropic. 

Based on the Debye model for anisotropic systems, proposed by Z. Chen, \textit{et al.} in Ref.\cite{Chen-PRB.87.125426(13)}, we adapted the single mode relaxation time model in combination with a first order time-dependent perturbation to the anharmonic ionic potential \cite{Srivastava-Book(90), Cuffe-PRL110-9(13)} to calculate the observed phonon damping time as function of its frequency.\\

\begin{figure}[t]
\includegraphics*[keepaspectratio=true, clip=true, angle=0, width=.5\columnwidth, trim={0mm, 0mm, 0mm, 0mm}]{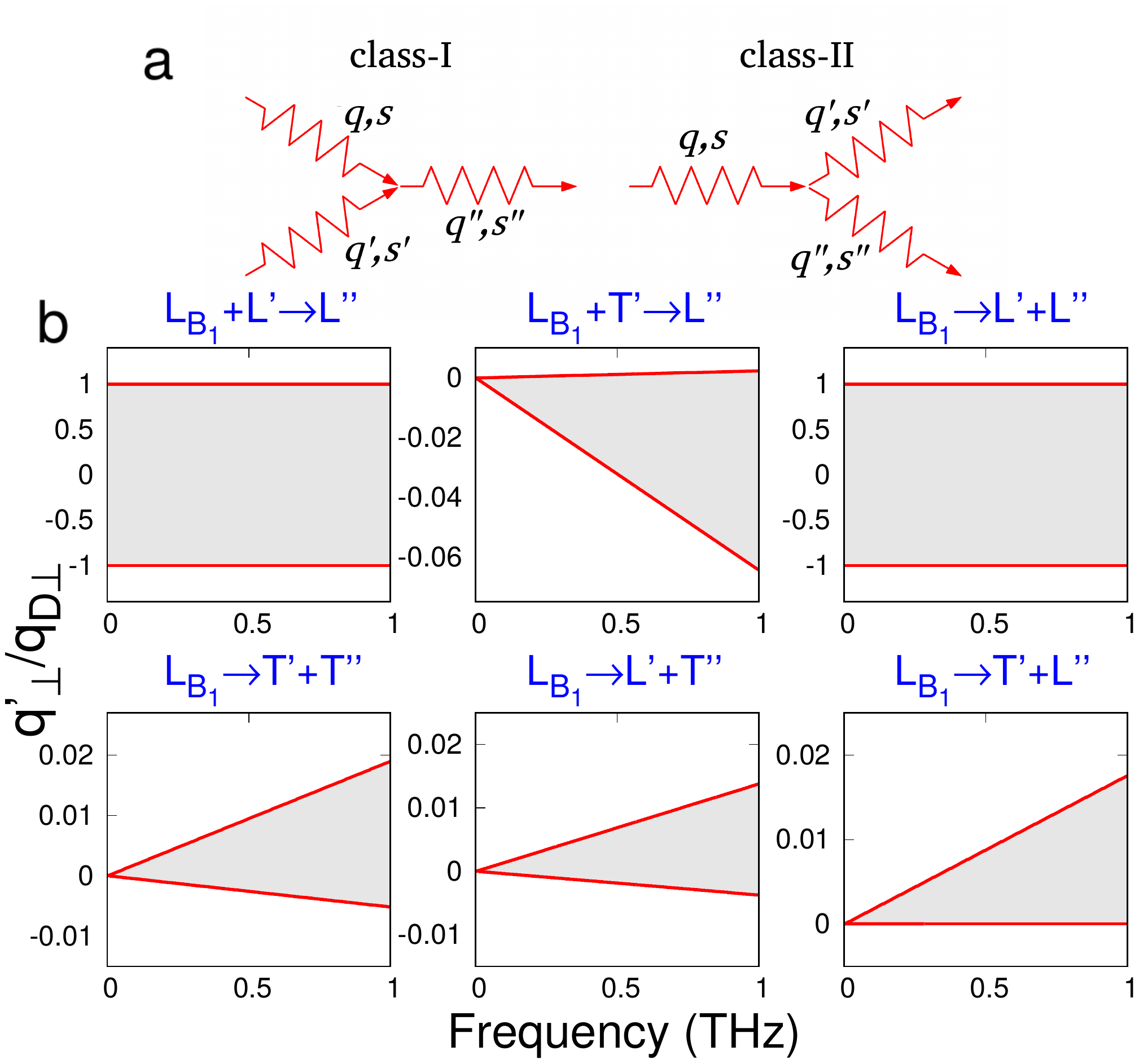}
\caption{\textbf{a)} Feynman diagrams for the class-I and class-II processes. \textbf{b)} areas that define the integration limits for eqn.\eqref{tau3}, for the variable $q'_{\perp}/q_D$  as function of the $L_{B_1}$ phonon frequency, for the different three-phonon processes.}
\label{figura:SM8}
\end{figure}

The initial state of the phonon system is described by $\vert i\rangle = \vert n_{qs}, n_{q's'}, n_{q''s''}\rangle$, where $q$ and $s$ identify the wave vectors and polarizations of each phonon mode, respectively, and $n_{qs}$ indicates the number of phonon in the state $q,s$. The anharmonic perturbation $H_{anh}$ causes the system to scatter to a final state. Within a three-phonon process, two possibilities should be taken into account: First, the combination of two initial phonons to a third one (\textit{class-I} process), i.e. $\omega_{q,s}+\omega_{q',s'}\rightarrow\omega_{q'',s''}$, case in which the final state is $\vert f\rangle = \vert n_{qs}-1, n_{q's'}-1, n_{q''s''}+1\rangle$. And second, the annihilation of a phonon into two remaining (\textit{class-II} process), i.e. $\omega_{q,s}\rightarrow\omega_{q',s'}+\omega_{q'',s''}$, where the final state is of the form $\vert f\rangle = \vert n_{qs}-1, n_{q's'}+1, n_{q''s''}+1\rangle$ \cite{Srivastava-Book(90), Maris-PhysicalAcoustics-Book(71)}. Figure \ref{figura:SM8}a shows the Feynman diagrams for both processes. The rate of occurrence $P_{3ph}$ of them, is given by Fermi's golden rule \cite{Srivastava-Book(90)}
\begin{eqnarray}
P_{3ph} = \frac{2\pi}{\hslash} \left| \langle f \vert H_{anh} \vert i \rangle \right|^2 \delta(E_f-E_i ),
\end{eqnarray}
where the delta function ensures the conservation of energy between initial and final state. \\

Since our interest is centred at a state with an initial phonon with wavevector $q$ and polarization state $s$ (the $B_1$ longitudinal acoustic mode), its total single mode relaxation rate is proportional to $P_{3ph}$, and is given by \cite{Srivastava-Book(90)}
\begin{eqnarray}\label{tau}
\begin{split}
\tau_{qs}^{-1} = \frac{\pi\hslash}{4\rho^3N_0\Omega}  \sum_{q's'q''s''} \big|A_{qq'q''}^{ss's''}\big|^2 \mbox{$\frac{qq'q''}{v_sv_{s'}v_{s''}}$}\delta_{q+q'+q'',G}
\left[ \frac{\bar{n}_{q's'}(\bar{n}_{q''s''}+1)}{\bar{n}_{qs}+1}\delta(\omega_{qs}+\omega_{q's'}-\omega_{q''s''}) +\right.\\ \left.\mbox{$\frac{1}{2}$} \frac{\bar{n}_{q's'}\bar{n}_{q''s''}}{\bar{n}_{qs}}\delta(\omega_{qs}-\omega_{q's'}-\omega_{q''s''})\right] ,
\end{split}
\end{eqnarray}
where the first term within brackets corresponds to transitions of class-I, while the second term to transitions of class-II. $\bar{n}_{qs}=[\exp(\hbar \omega_{qs}/k_BT)-1]^{-1}$ are the Bose-Einstein distribution functions for phonons in the state $q$,$s$ and temperature $T$, $G$ is a vector of the reciprocal lattice, $N_0$ the number of unit cells of volume $\Omega$ and density mass $\rho$, and the factor $\left|A_{qq'q''}^{ss's''}\right|^2$ is the three-phonon scattering strength of the anharmonic perturbation ($H_{anh}$) for the three-phonon process. Under some assumptions this strength can be approximated as a $q$ independent magnitude, as\,\cite{Srivastava-Book(90)}
\begin{eqnarray}\label{A}
\big|A_{qq'q''}^{ss's''}\big|^2 = \frac{4\rho^2}{\bar{v}^2}\gamma^2v_s^2v_{s'}^2v_{s''}^2.
\end{eqnarray}
Here $\gamma$ is the mode independent Gr\"{u}neisen constant, $\bar{v}^2$ the phonon average group velocity and $v_{s}$, $v_{s'}$ and $v_{s''}$ the corresponding phonon group velocities for each of the involved acoustic phonon modes.\\

To evaluate eqn.\eqref{tau} it is necessary to sum over all different phonon states $q',s'$ and $q'',s''$ within the Brillouin zone. This involves the knowledge of the full acoustic phonon dispersion relation. In order to get more insight and understand the phenomenology underlying the involved processes, we follow the well known Debye approximation that states a linear dispersion relation for transverse and longitudinal phonon modes \cite{Srivastava-Book(90), Cuffe-PRL110-9(13)}. It is important to mention that the Debye approximation is an approach that accounts for a strictly continuous Brillouin zone, and the modes in the flake's out-of-plane direction are in fact discrete. Consequently, it is expected to have a rather closer agreement with this model for thicker samples, i.e. where the acoustic free spectral range of the flake in the perpendicular direction is smaller, and for lower frequencies. Clearly, these kind of model excludes acoustic shear modes such as those present in atomically thin layered materials (e.g. Z-modes with parabolic dispersion relation). Being aware of these limitations, we intend to obtain a rather qualitative description that allows to get a better understanding of the processes involved in the decay of the coherently generated acoustic modes. \\

In an attempt to capture the essence of these kind of layered materials, we introduce the anisotropic Debye approximation \cite{Chen-PRB.87.125426(13)}. A layered system such as the one we analyse in this work, has an important anisotropy, which is evidenced by the different velocity of sound in the in-plane direction ($v_{\parallel}$) and in the stacking (out-of-plane) direction ($v_{\perp}$). Neglecting the in-plane differences, the model proposes an ellipsoidal iso-energy as function of the wavevector $q$ of the form  
\begin{eqnarray}\label{anisotropic disp rel Debye}
\omega^2_{q,s} = v_{\parallel,s}^2 q_{\parallel,s}^2 + v_{\perp,s}^2 q_{\perp,s}^2 ,
\end{eqnarray}
where $\omega$ is the angular frequency of the acoustic phonon, $q_{\parallel}$($q_{\perp}$) the wavevector in the in-plane(out-of-plane) direction and $v_{\parallel,s}$ should be taken as an average of the in plane sound velocities for the polarization $s$.\\

Considering the relation given by eqn.\eqref{A}, replacing the sum over $q''$ evaluating the Kr\"onecker delta function, changing the sum over $q'$ in eqn.\eqref{tau} to $\sum_{q'}\rightarrow\frac{N_o\Omega}{8\pi^3}\int d^3q'$, expressing the integral in cylindrical coordinates $d^3q'\rightarrow q'_{\parallel}dq'_{\parallel}d\varphi_{q'}dq'_{\perp}$, and the anisotropic dispersion relation given by eqn.\eqref{anisotropic disp rel Debye}, we get
\begin{eqnarray}\label{tau2}
\begin{split}
\tau_{qs}^{-1} = \frac{\hslash \gamma^2}{4\pi\rho \bar{v}^2}  \sum_{s's''G} \int v_sv_{s'}v_{s''}q_{B_1}q'q''
 \left[ \frac{n_{q's'}(n_{q''s''}+1)}{n_{q_{B_1}s}+1}\delta(\omega_{q_{B_1}s}+\omega_{q's'}-\omega_{q''s''}) +\right.\\ \left.\mbox{$\frac{1}{2}$}\frac{n_{q's'}n_{q''s''}}{n_{q_{B_1}s}}\delta(\omega_{q_{B_1}s}-\omega_{q's'}-\omega_{q''s''})\right] q'_{\parallel} dq'_{\parallel}dq'_{\perp}\ ,
\end{split}
\end{eqnarray}\\
where $q$ in eqn.\eqref{tau} was replaced by $q_{B_1}$, since it corresponds to the generated and observed $B_1$ mode that is a longitudinal acoustic mode ($L_{B_1}$) along the $z$ direction.

The conservation of the momentum leads then to 
\begin{eqnarray}\label{q''}
q'' = \sqrt{q{'}_{\parallel}^2 + (q_{B_1} \pm q'_{\perp})^2},
\end{eqnarray}
for the normal ($G=0$) class-I (+) and class-II (-) processes. The Umklapp processes ($G \neq 0$) are discarded for reasons that will be explained later.\\

The integral over $q'_{\parallel}$ in eqn.\eqref{tau2} can be further evaluated by using the Dirac delta function through the substitution
\begin{eqnarray}\label{Delta}
\Delta = \omega_{q_{B_1}s}\pm\omega_{q's'}-\omega_{q''s''},
\end{eqnarray}
leading to
\begin{eqnarray}\label{DDelta}
d\Delta = \left(\pm\frac{d\omega_{q's'}}{dq'_{\parallel}}-\frac{d\omega_{q''s''}}{dq'_{\parallel}.}\right)dq'_{\parallel}
\end{eqnarray}
$\omega_{q''s''}$ depends on $q'_{\parallel}$ through the relation given by eqn.\eqref{q''}. Expression \eqref{tau2} reads then
\begin{eqnarray}\label{tau3}
\begin{split}
\tau_{qs}^{-1} = \frac{\hslash \gamma^2}{4\pi\rho \bar{v}^2} \sum_{s's''} \int v_sv_{s'}v_{s''}q_{B_1}q'q''
 \left[\frac{n_{q's'}(n_{q''s''}+1)}{n_{q_{B_1}s}+1} \left. \frac{\omega_{q's'} \omega_{q''s''}}{{v'_{\parallel}}^2\omega_{q''s''} - {v''_{\parallel}}^2\omega_{q's'}}\right|_{q'_{\parallel 0}}\right. -\\  \left. \left.\frac{1}{2}\frac{n_{q's'}n_{q''s''}}{n_{q_{B_1}s}}\frac{\omega_{q's'} \omega_{q''s''}}{{v'_{\parallel}}^2\omega_{q''s''} + {v''_{\parallel}}^2\omega_{q's'}}\right|_{q'_{\parallel 0}} \right]  dq'_{\perp},
\end{split}
\end{eqnarray}
where $q'_{\parallel 0}$ is the value of $q'_{\parallel}$ that makes the argument of the Dirac delta functions in eqn.(\ref{tau2}) zero, i.e. $\Delta = 0$. The explicit expression for $q'_{\parallel 0}$ depends on the class of process and the polarizations of the involved phonon modes.\\

It remains to be evaluated the limits of the integral \eqref{tau3} in $q'_{\perp}$. As the Debye approximation implies a cut-off frequency $\omega_D$ for acoustic waves in a crystal, in an anisotropic material it is possible to define the characteristic Debye frequencies of the plane ($\omega_{D\parallel}$) and perpendicular to the plane ($\omega_{D\perp}$) as \cite{Chen-PRB.87.125426(13)},
\begin{eqnarray}\label{w_D}
\omega_{D\parallel} = v_{\parallel} q_{D\parallel}, \quad \omega_{D\perp} = v_{\perp} q_{D\perp},
\end{eqnarray}
where $q_{D\parallel}$($q_{D\perp}$) is the in-plane (perpendicular to the plane) cut-off wavevector. The ellipsoid
\begin{eqnarray}\label{q_cutoff}
\frac{q_{\parallel}^2}{q_{D\parallel}^2} + \frac{q_{\perp}^2}{q_{D\perp}^2} = 1
\end{eqnarray}
defines the cut-off wavevector for directions that are neither parallel nor perpendicular to the plane. The number density of the primitive cell ($\eta$) relates the total number of acoustic modes with the cut-off wavevectors as \cite{Chen-PRB.87.125426(13)}
\begin{eqnarray}\label{eta}
\eta = 3n_0/\Omega = \frac{1}{6\pi^2}q_{D\parallel}^2q_{D\perp},
\end{eqnarray}
where $n_0$ is the number of effective masses in the unit cell and $\Omega$ its volume. As explained in section \ref{sec-LCM}, $n_0 = 2$. The cut-off wavevector ellipsoid \eqref{q_cutoff} is completely defined through \eqref{eta} and the anisotropy ratio $q_{D\parallel}/q_{D\perp}$, that can be approximated by the extents of the first Brillouin zone in the corresponding directions \cite{Chen-PRB.87.125426(13)}, 
\begin{eqnarray}\label{ratio}
\frac{q_{D\parallel}}{q_{D\perp}} = \frac{c}{a},
\end{eqnarray}
where $c$ ($a$) is the lattice parameter in the direction perpendicular (parallel) to the layers.

The limits in the integral \eqref{tau3} are determined by the possible real values of $q'_{\parallel}$ and $q'_{\perp}$ defined by the Dirac delta function ($\Delta = 0$) and by imposing that $q'$ belongs to the cut-off wavevector ellipsoid.

From the different possible interactions giving raise to the $B_1$ mode's decay, only some of them need to be accounted for, namely: $L_{B_1}+L'\rightarrow L''$, $L_{B_1}+T'\longrightarrow L''$ for the class-I processes, and $L_{B_1}\longrightarrow L'+L''$, $L_{B_1}\longrightarrow T'+T''$, $L_{B_1}\longrightarrow L'+T''$, and $L_{B_1}\longrightarrow T'+L''$ for the class-II processes. $L'$, $L''$ ($T'$, $T''$) are the corresponding longitudinal (transversal) acoustic phonons involved in the three-phonon scattering processes.\\

Figure \ref{figura:SM8}b shows the areas that define the integration limits for the variable $q'_{\perp}$ in eqn.\eqref{tau3} normalized to $q_{D\perp}$. For a given $B_1$ mode frequency, the integration in $q'_{\perp}$ is constrained depending on the different possible interactions. Only the $L_{B_1}+L'\rightarrow L''$, and $L_{B_1}\longrightarrow L'+L''$ processes are limited by the cut-off wavevector ellipsoid and, since they are collinear ($q'_{\parallel}=0$), the integration in $q'_{\perp}$ goes from -1 to 1. As these ranges of integration are much bigger than in the other processes, the main contribution to the phonon lifetime are due to them.

Different possible Umklapp processes that could be taken into account are: $L_{B_1}+T'\longrightarrow L''$ for the class-I processes, and $L_{B_1}\longrightarrow T'+T''$, $L_{B_1}\longrightarrow L'+T''$, and $L_{B_1}\longrightarrow T'+L''$ for the class-II processes \cite{Srivastava-Book(90)}. However, the calculation shows that the Umklapp processes are not accessible since there are no $q'$ that fulfil the energy and momentum conservation for the involved $B_1$ mode frequencies, as it was observed in [\onlinecite{Cuffe-PRL110-9(13)}] for an isotropic material.

\begin{figure}[ttt]
\includegraphics*[keepaspectratio=true, clip=true, angle=0, width=.5\columnwidth, trim={0mm, 15mm, 0mm, 18mm}]{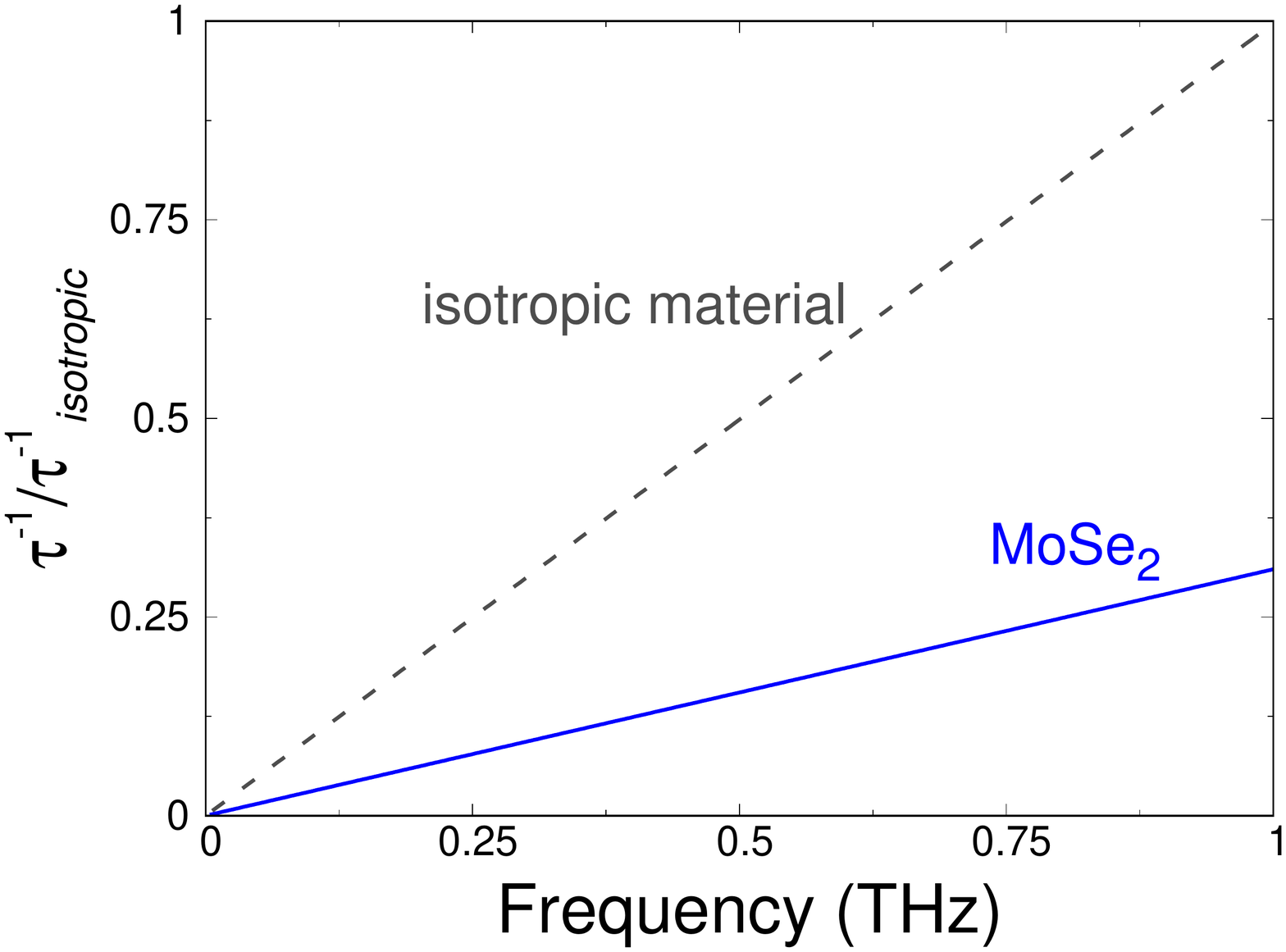}
\caption{Calculated phonon lifetime as a function of the frequency for MoSe$_2$, and the calculation for an equivalent isotropic material.}
\label{figura:SM9}
\end{figure}

For the calculations, the in-plane speeds of sound for bulk MoSe$_2$ were estimated from the simulated phonon dispersion relation from Ref.[\onlinecite{Kumar-ChmOfMaterials27-4-1278(15)}], where we derived $v_{\parallel L} \simeq 5800$\,m/s and $v_{\parallel T} \simeq 2900$\,m/s for the longitudinal and transversal velocities, respectively. For the out-of-plane components, we used the velocities estimated in the main text and in section \ref{sec-LCM}. We perform the explicit evaluation of the integral in eqn.\eqref{tau3}, including the processes sketched in Fig.\ref{figura:SM8}. The result for the inverse $B_1$ modes lifetime ($\tau^{-1}$) for MoSe$_2$ is shown in Fig.\ref{figura:SM9} (blue lines), and also in Fig.\,4 of the main text. For the latter the proportionality factor was fitted to match the experimental data. Since many of the parameters involved are unknown (Gr\"uneisen parameter, three-phonon scattering strength, etc.), $\tau^{-1}$ in Fig.\ref{figura:SM8} is shown relative to the value obtained for an equivalent isotropic material, where the isotropic $L$ and $T$ speeds of sound are calculated as the average of those used for the MoSe$_2$: $\bar{v}_L = \frac{2}{3} v_{\parallel L}+ \frac{1}{3} v_{\perp L} = 4800$\,m/s and $\bar{v}_T = \frac{2}{3} v_{\parallel T}+ \frac{1}{3} v_{\perp T} =2450$\,m/s. An important point to be mentioned is, that both calculated curves (anisotropic and isotropic) result \emph{linear} with the frequency ($\tau^{-1} \propto f$). As many of the parameters included in the proportionality factor are unknown, it is not possible to compare both calculations with the measurements. It follows from Fig.\ref{figura:SM9} that the anisotropic simulation gives scattering rates that are significantly lower than the isotropic equivalent, i.e. the lifetimes for the anisotropic system is increased with respect to the equivalent isotropic system. The reason for this is that, as explained before, the mayor contributions to the lifetime comes from the collinear processes ($q'_{\parallel}=0$).  Its magnitude depends on the cut-off wavevector in this direction: $q_{D\parallel}$ for the anisotropic case, and $q_D$ for the isotropic case. Since $q_D < q_{D\parallel}$, the lifetime is consequently increased for the anisotropic case. 


\section{Surface roughness mechanism for the decay of phonons}\label{sup}

\begin{figure}[ttt]
\includegraphics*[keepaspectratio=true, clip=true, angle=0, width=.5\columnwidth, trim={0mm, 15mm, 0mm, 15mm}]{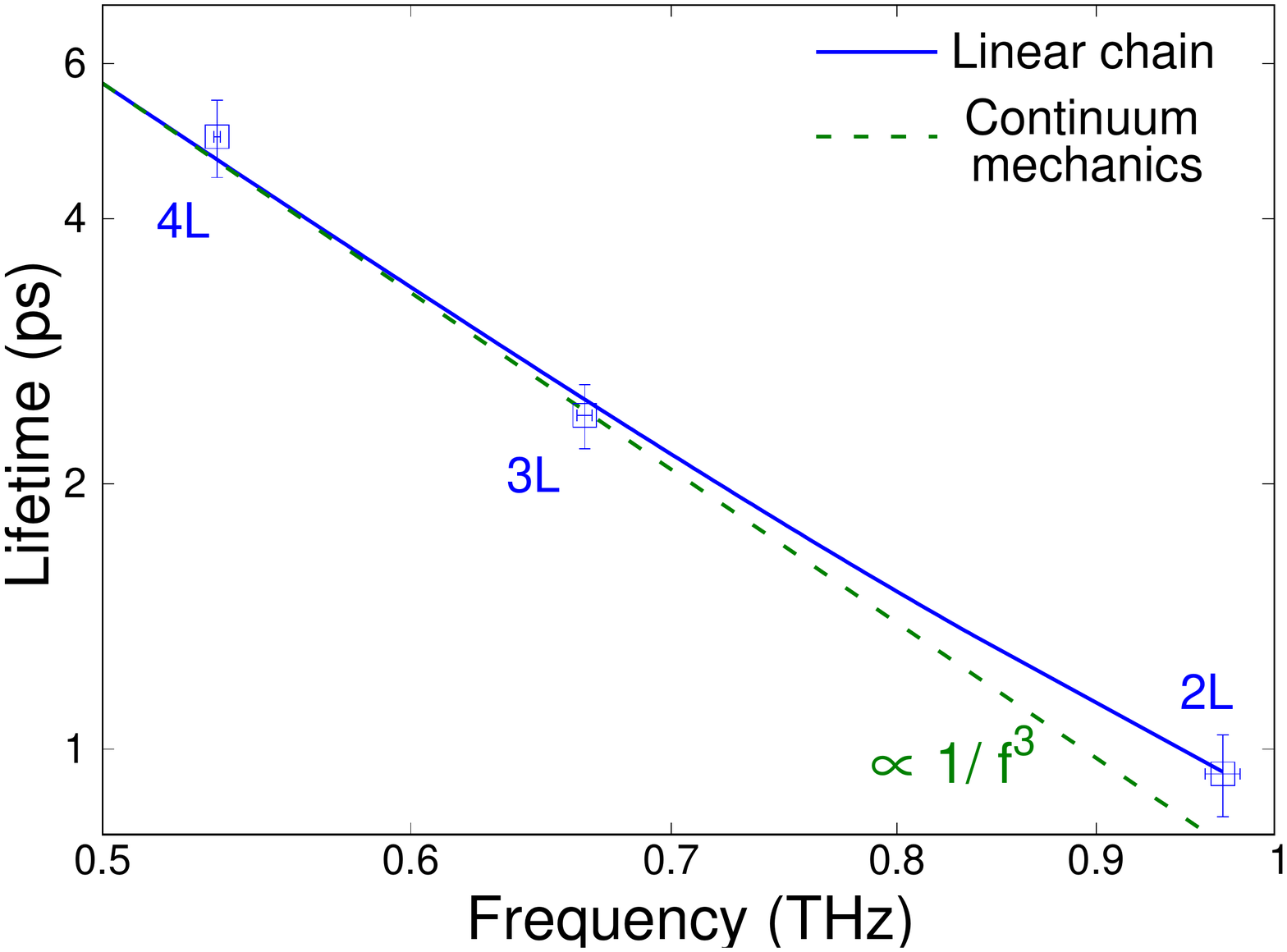}
\caption{Calculated relaxation time due to the boundary scattering for the linear chain model (blue) and the continuum mechanics model (dashed green line). The experimental data is shown with blue squares.}
\label{figura:SM-rough}
\end{figure}
The effect of the roughness on the phonon lifetime becomes more important when reducing the flake's thickness. In order to model this effect, we follow the approach of Ziman [\onlinecite{Ziman-Book(60)}], described in the main text, that takes into account the surface \textit{asperity} $\eta$ and the associated imperfect (non specular) reflection of the acoustic wave at the flake's surfaces. The relaxation time due to the boundary scattering effects takes the form \cite{Ziman-Book(60), Cuffe-PRL110-9(13)}
\begin{eqnarray}\label{sup1}
\tau_b = \frac{Nd_0}{v_s}\frac{1+p}{1-p},
\end{eqnarray}
where $p(\lambda_{ac}) = \exp[-16\pi^2\eta^2/\lambda_{ac}^2]$ is the wavelength dependent specularity \cite{Ziman-Book(60), Cuffe-PRL110-9(13)}. 

Reordering (\ref{sup1}), it is possible to express this contribution to the phonon lifetime as
\begin{eqnarray}\label{sup2}
\tau_b = \frac{\lambda_{ac}}{2v_s}\coth\left(\frac{8\pi^2\eta^2}{\lambda_{ac}^2}\right).
\end{eqnarray}
Within a continuum elastic limit, the acoustic pulse follows a linear dispersion relation. Under this approximation, as mentioned in the main text $v_s\rightarrow v_{ac}$, and the wavelength ($\lambda_{ac}$) of the acoustic phonons relates to its frequency as $\lambda_{ac}=v_{ac}/f$. Replacing this relation into eqn.\eqref{sup2} and assuming that the argument of the hyperbolic cotangent is small enough, the first order Taylor expansion yields
\begin{eqnarray}\label{sup3}
\tau_b \simeq \frac{v_{ac}^2}{16\pi^2\eta^2}f^{-3}.
\end{eqnarray}
Note that under these circumstances, the boundary contribution $\tau_{b}$ results proportional to $f^{-3}$. This rough approximation is used by some authors \cite{Cuffe-PRL110-9(13)}. However, it is not entirely correct for our case, since the argument of the hyperbolic cotangent ($8\pi^2\eta^2/\lambda_{ac}^2$) is not small enough for the frequency region corresponding to the thinnest samples. In Figure \ref{figura:SM-rough}, the comparison between the calculation of $\tau_b$ with eqn. (\ref{sup2}) and using (\ref{sup3}) is shown, displaying the range where the difference between both expressions is more notorious. The continuous line is the one used in Fig.4 of the main text, and the dashed lines corresponds to the approximated expression. The blue squares are the experimental data, that correspond to 2L, 3L and 4L-MoSe$_2$. As observed the difference is small.

\section{Acoustic quality factor}\label{6}

\begin{figure}[h]
\includegraphics*[keepaspectratio=true, clip=true, angle=0, width=.5\columnwidth, trim={0mm, 14mm, 0mm, 18mm}]{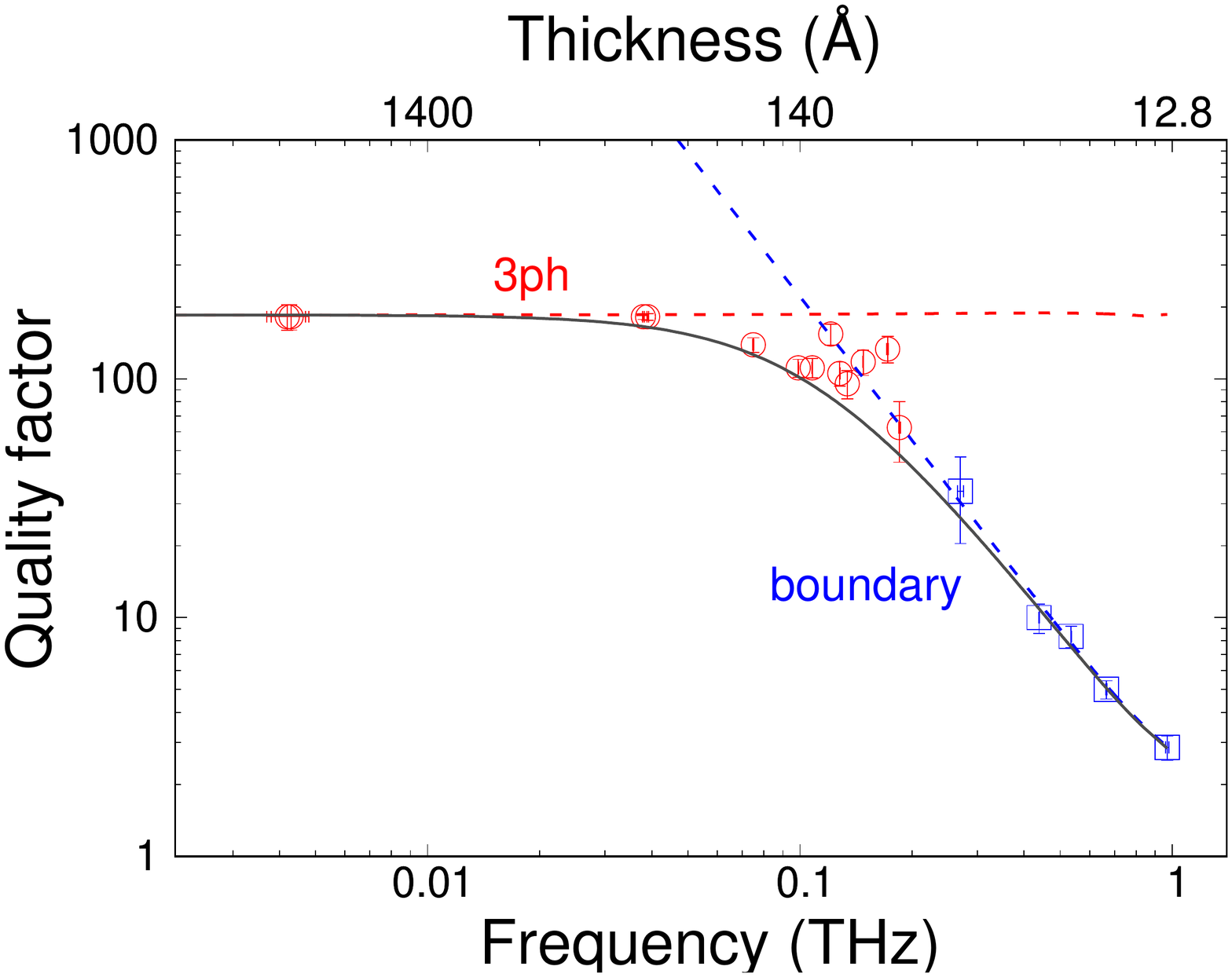}
\caption{Acoustic quality factor $\mathcal{Q}$ as a function of the $B_1$-mode frequency. The color coding is the same as the one in Fig.2 and 4 of the main text.}
\label{figura:SM7}
\end{figure}

A parameter that is of interest for applications, e.g. for the optomechanics community is the quality factor ($\mathcal{Q}$-factor) of the system considered as a mechanical oscillator. The $\mathcal{Q}$-factor of such an oscillator, given the measured lifetimes as a function of the mode frequency, can be obtained by the simple expression:
\begin{eqnarray}\label{eqn:Q}
\mathcal{Q}=\pi f \tau.
\end{eqnarray}
Figure \ref{figura:SM7} presents this magnitude for the measured membranes as a function of the frequency $f$ of the $B_1$-mode calculated from the obtained modes' lifetime $\tau_{B_1}$ (see Fig.\,4 in the main text), and eqn.\eqref{eqn:Q}. The colors for the experimental points and simulated lines coincide with those in Fig.\,2 and 4 of the main text. The constant behaviour for lower frequencies results from the linear dependence with $1/f$ of the phonon lifetime in this frequency range, that cancels out with the multiplying $f$ in eqn.\eqref{eqn:Q}. For the thinner samples, in which the phonon decay is dominated by the boundary scattering and the phonon lifetime becomes $\tau \propto 1/f^3$, the $\mathcal{Q}$-factor is progressively reduced. Again, by means of Matthiensen's rule \cite{Ziman-Book(60)}, the quality factor in the entire range is plotted with the grey dashed line. 

The $\mathcal{Q}$-factor obtained in this work, considering the range where the phonon lifetime depends as $1/f$ with the frequency, is of the order of those obtained for single crystalline silicon membranes in Refs.[\onlinecite{Bruchhausen-PRL106-077401(11)}] and [\onlinecite{Cuffe-PRL110-9(13)}].




\end{document}